# Mitigating business risks from renewable PPA power sourcing uncertainties for European green hydrogen production: Robust system design, regulatory adjustments and offtake flexibility

J.Brandt*, A.Bensmann*, R. Hanke-Rauschenbach*
* Leibniz Universität Hannover, Institute of Electric Power Systems, 30167 Hanover, Germany

Brandt, Jonathan (Conceptualization, Software, Analysis, Visualization, Methodology, Writing - original draft)

Bensmann, Astrid; Hanke-Rauschenbach, Richard (Conceptualization, Project administration, Supervision, Writing - Review & editing, Funding acquisition)

## Broader context

Green hydrogen is a pivotal cornerstone of decarbonising the European Union's industry, transport and power sectors, and simultaneously breaking free from energy dependencies amidst war-driven energy price explosions. However, the ramp-up of local green hydrogen production falls short of expectations, among others, hampered by overly strict rules on renewable power purchase for water electrolyser operation when producing green hydrogen, prompting Member states' ministries in early 2026 to urge the European Commission to advance a planned rules review from 2028 to 2026. By taking the perspective of an electrolyser operator, we show how the interaction between key power purchase rules can jeopardise the fulfilment of green hydrogen offtake agreements, affect production costs and thus cause delayed investment decisions and cancelled projects. Furthermore, we evaluate different measures regarding production system design, renewable power purchase, green hydrogen offtake flexibility and regulatory adjustments regarding their effectiveness in risk mitigation. The findings provide valuable insights for important economic players along the green hydrogen supply line and help regulators broaden their understanding in the upcoming review of power sourcing rules by addressing an important gap in the existing literature on the effects of power sourcing rules on green hydrogen production.

## Abstract

As energy prices surge for the second time in recent years driven by the ongoing crisis in the Middle East, the European Union's continuing reliance on fossil energy imports is becoming increasingly apparent. However, despite offering an intriguing prospect of improved energy resilience, the ramp-up of local green hydrogen production lags far behind the officially stated ambitions set after the 2022 energy crisis. A prominent reason for the widening implementation gap between announced and realised production projects is overly strict rules on renewable power sourcing, prompting Member states' ministries and the European Commission to propose advancing a planned rules review from 2028 to 2026. To contribute to a successful review and rule adjustments, we address an important gap in understanding the effects of power purchase rules on green hydrogen production. By taking the perspective of European electrolyser operators, we show how the criterion of additionality and its interaction with required temporal correlation can jeopardise the fulfilment of green hydrogen offtake agreements and affect green hydrogen production costs across different European bidding zones. Applying different design paradigms to a green hydrogen production system reveals that electrolyser operator measures, such as PPA and storage upsizing, can help to mitigate the business risks posed by the additionality criterion but come with increased costs. Alternatively, relaxed temporal correlation and increased offtake flexibility both increase production system robustness and reduce production costs simultaneously. Whereby relaxing temporal correlation rules does not result in exceeded emission intensity thresholds, underlining the potential of extended transitional rules to support the ramp-up of European green hydrogen production.

# 1 Introduction

The subject of this article are the European Union power sourcing rules for renewable power purchase agreements (PPAs) used for green hydrogen production, their effect on the fulfilment of green hydrogen offtake agreements, and different business risk mitigation measures.

Following an episode of enthusiasm surrounding the opportunities of renewable (henceforth "green") hydrogen to achieve European emission reduction goals [1] and to reduce dependence on energy imports [2], the implementation gap between announced and realised projects mirrors the slowed ramp-up of European green hydrogen production in recent years [3]–[5]. Reasons include limited and uncertain green hydrogen offtake [4]–[6] and a lack of willingness to pay the high, uncertain production and supply costs compared to hydrogen from fossil sources [3], [7]. Besides other factors, the high costs are driven by slower-than-expected declines in capital expenditures [8] and regulatory requirements for power sourcing for electrolytic green hydrogen production [9]–[11]. Both impair production profitability and thus hamper a European ramp-up, necessitating high project subsidies [3], [5], [12] to mitigate business risks. However, in many cases even heavily subsidised projects lack economic viability and thus are not implemented [4], [5], [13]. Consequently, in the context of the ongoing crisis in the Middle East, revealing Europe's remaining dependency on fossil energy imports, ministries from across the European Union jointly called on the European Commission to bring forward the review of the power sourcing requirements for green hydrogen production planned for 2028, and to finalise it in 2026 [5]. With the introduction of the Communication *AccelerateEU* in April 2026 [14], which immediately addressed the call for a preliminary review, the importance of a comprehensive understanding of the impact of the regulatory framework on European green hydrogen production is rapidly growing.

The European Union's regulatory framework requires the primary use of renewable power via PPAs for green hydrogen production in most European bidding zones in upcoming years [9], [15], leading to multiple challenges for electrolyser operators once power sourcing is on the agenda. The use of renewable PPAs for green hydrogen production is subject to three criteria, which shall ensure emission savings compared to fossil comparators. First, the geographical correlation criterion states that the electrolyser and the renewable PPAs shall be situated within the same bidding zone. Second, the simultaneity of hydrogen production and renewable PPA power feed-in shall be ensured by an hourly correlation. This poses the challenge of matching the fluctuating feed-in from renewable energy sources to the offtaker requirements of European green hydrogen consumers in the steel-, refinery-, chemical- and fertiliser-industries [16]. Third, the additionality criterion for renewable PPAs leads to an exclusive reliance on newly built renewable power plants commissioned within the last 36 months before the electrolyser. By excluding existing and depreciated renewable power sources, this criterion limits the availability of bookable PPAs to only newly built, presumably more expensive options. In addition, the newly built PPAs are mainly available through long-term contracts, limiting power sourcing flexibility and increasing power supply risks due to the inter-year uncertainties in weather-reliant renewable PPA power feed-in. This applies to all countries where short-term contracts via existing renewable power plants are not allowed for green hydrogen production, and as long as short-term power purchase via the electricity market for compensation is not an option due to the strict temporal correlation criteria, as is the case in most European bidding zones from 2030 onward [9], [15]. Here, someone along the supply chain, the PPA-operator, the electrolyser-operator or the green hydrogen offtaker, must bear the resulting inter-year volume risk.

## 1.1 Recent literature discussed

Recent literature provides different analyses of the power sourcing criteria for green hydrogen production and their effects on production costs and associated emissions. At the energy system level, three prominent studies evaluate the interacting effects between the temporal correlation and additionality criteria [17]–[19]. Zeyen et al. [17] show that additional local renewables are mandatory to avoid increased emissions in green hydrogen production in central Europe. Furthermore, their results emphasise that following strict temporal correlation rules to avoid increased emissions does not necessarily have to lead to higher production costs. Namazifard et al. [18] identify the additionality criterion to be the dominant cost driver for green hydrogen production across the European Union and emphasize the necessity for country-specific exemptions from the power sourcing criteria. Taking a more international analysis perspective, the findings of Giovanniello et al. [19] underline that the system emissions from different temporal correlation rules are highly dependent on the interpretation of the additionality criterion.

Complementary to system-level research, multiple studies take the perspective of electrolyser operators to evaluate the effects of the European power purchase rules on economic and emission-related aspects [9], [11], [20]–[22]. The first study on this topic takes the perspective of an electrolyser operator in Germany. It focuses on the effects of the temporal correlation criterion on the cost of green hydrogen production and emissions [11]. Brandt et al. [9] take a similar approach but widens the picture of possible effects across Europe by evaluating the impact of different uncertain boundary conditions, like the availability of renewable energy and grid electricity prices. In this context, Ferrùs et al. [20] provide insights on how an optimised PPA portfolio of mixed locations and technologies can substantially reduce production costs for electrolyser operators in Germany when strict temporal correlation conditions apply. Also important to mention in this regard are two studies that take a similar perspective but have a stronger focus on the offtake side and use stochastic rather than of deterministic methods [21], [22]. Brucksch et al. [21] use stochastic PPA portfolio optimisation to demonstrate how different temporal correlation rules and PPA diversification affect production costs and emissions of a green hydrogen production project providing for a glasswork industry in Germany. Whereas Palmer et al. [22] evaluate the suitability of different deterministic and stochastic planning methods for green hydrogen production projects in France for addressing uncertainties along the supply chain, with a focus on different hydrogen offtake contracts and taking temporal correlation rules into account.

## 1.2 Contribution of present article

In summary, despite a more comprehensive coverage of power purchase criteria at the system level, research lacks insights into how the additionality criterion affects green hydrogen production from an electrolyser operator's perspective and how the different criteria interact in this context. With the potential preliminary review of the power purchase criteria in 2026, providing the most comprehensive understanding possible of their effects on green hydrogen production is crucial to enable the European Commission to make well-considered adjustments. Emphasised by the joint call for a preliminary review by several ministries of European Union member states [5], suitable adjustments can accelerate the slowed ramp-up of European green hydrogen production, thereby closing the widening implementation gap, simultaneously increasing Europe's energy independence, and ensuring emission reductions. To contribute in this regard, this study explores how the criterion for additional renewable PPAs and the resulting long-term contracts, along with their interactions with temporal correlation rules, affect green hydrogen production costs and the fulfilment of offtake agreements. In addition, it provides insights into how different electrolyser operator, offtaker, and regulatory measures could help to mitigate the resulting volume risk along the green hydrogen supply chain.

Therefore, a linear optimisation model of a green hydrogen production system comprising different power purchase options, an electrolyser plant and a hydrogen storage option, serves as the methodological basis for the analyses in this study and is introduced in Section 2.1. In Section 2.2, three different design paradigms are presented, which, when applied to the hydrogen production system, enable quantifying the impact of the additionality criterion on green hydrogen supply cost and the fulfilment of offtake agreements. Furthermore, their comparison provides insights into how operator measures regarding electrolyser-, storage-, and PPA-design can increase the robustness of renewable PPA-based green hydrogen production. In Section 2.3, different modifications to the hydrogen production system are introduced, the analysis of which provides a comprehensive picture of the effects. Furthermore, they give insights into how regulatory and offtaker measures could help mitigate volume risk along the supply line. In Section 3, we present the results of the different design paradigms and system modifications and put them into context. Subsequently, in Section 4 we elaborate on how the different electrolyser operator measures and increased offtake flexibility can help mitigate the volume risks along the green hydrogen supply line. In addition, we discuss whether and under what conditions regulatory adjustments to the power purchase rules for green hydrogen production, when reviewed in 2026, could increase the robustness of green hydrogen production projects.

# 2 Methodological approach

## 2.1 System under consideration

The considered hydrogen production system, which serves as the methodological basis for all analyses in this study, is shown in Figure 1 and consists of different power purchase options, a proton exchange membrane (PEM) electrolyser system, and a hydrogen storage option.

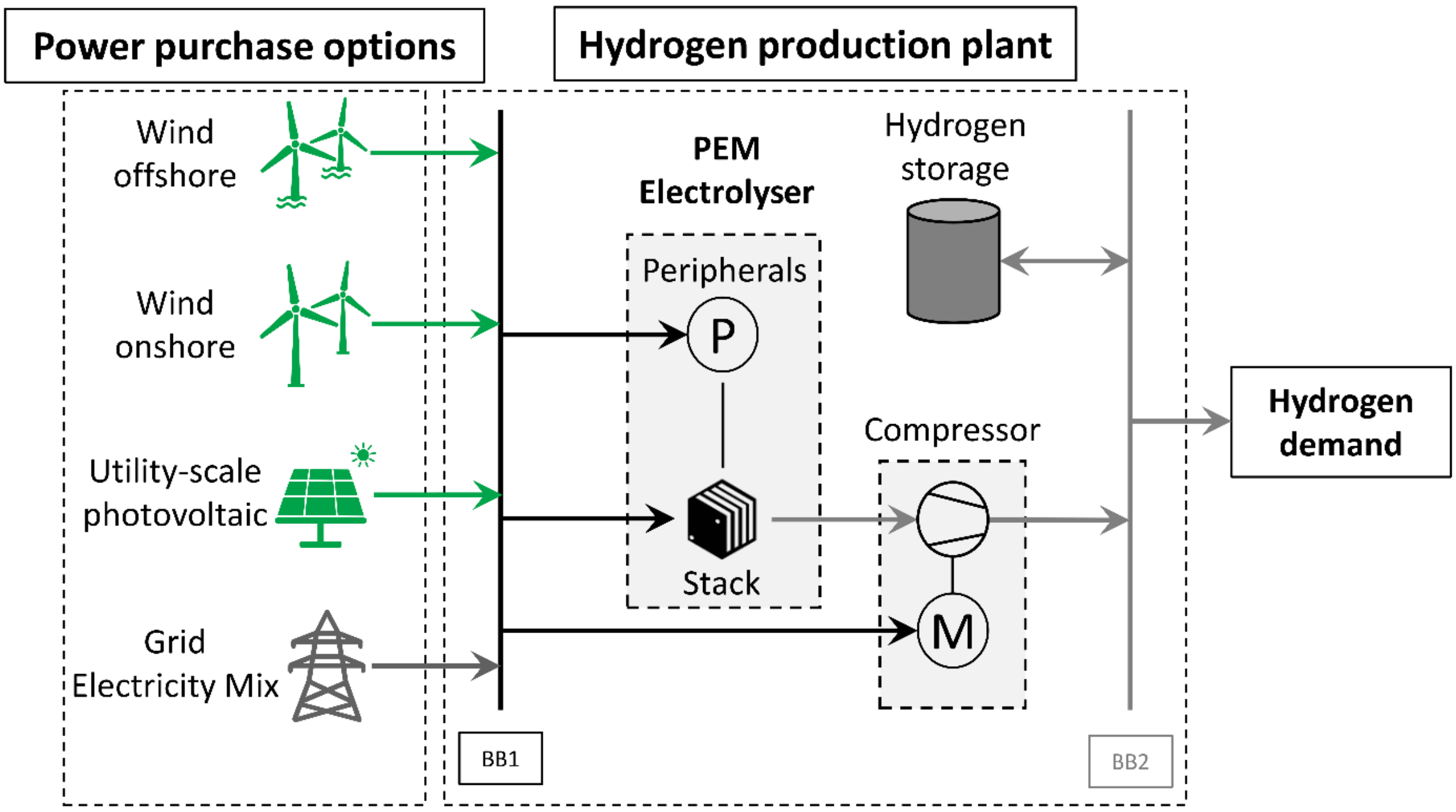

*Figure 1: Set-up of the modelled green hydrogen production system. The dashed box on the left shows all available renewable power purchase options. The dashed box on the right shows the hydrogen production plant consisting of a proton exchange membrane (PEM) electrolyser including peripherals (P) and stack, a piston compressor powered by an electric motor (M), and a hydrogen storage optionally as salt cavern storage bundle or pressure gas tank. Vertical lines: BB1, electricity bus bar; BB2, hydrogen bus bar.*

To be able to evaluate the economic performance of the considered system, its design and operation costs are minimised (see equation (1)) to meet a predefined green hydrogen demand.

$$\min \sum_{i=1}^{N} (C_{\mathrm{TAC,CAPEX},i} + C_{\mathrm{TAC,OPEX},i}) + \sum_{j=1}^{M} C_{\mathrm{TAC,PP},j} \tag{1}$$

The sum of the annualised installation cost $C_{\mathrm{TAC,CAPEX},i}$ and the annual operating cost $C_{\mathrm{TAC,OPEX},i}$ of all components *N* of the hydrogen production plant, and the annual power purchase cost $C_{\mathrm{TAC,PP},j}$ of all power purchase options *M,* make up the minimised cost. The onsite hydrogen supply costs (*OHSC*) are calculated by dividing the minimised annual cost by the annual sum of the predefined hydrogen demand (see equation (2)). They serve as an index for the economic performance of the considered system in the analyses performed in this study.

$$OHSC = \frac{\sum_{i=1}^{N} C_{\mathrm{TAC,CAPEX},i} + C_{\mathrm{TAC,OPEX},i} + \sum_{j=1}^{M} C_{TAC,PP,\mathrm{j}}}{\sum_{t=1}^{T} \dot{m}_{\mathrm{demand},t} \cdot \Delta t} \tag{2}$$

The renewable power purchase agreement options are wind offshore, wind onshore, and utility-scale photovoltaic, all of which are pay-as-produced. Supplementary Note 1 in the supplementary data contains further details on PPA modelling, including information on the used capacity factor time series and PPA price calculation. In the analyses with relaxed temporal correlation rules (see Section 3.4), where grid electricity mix is available as an additional power purchase option, fixed, average prices are assumed, and the pricing magnitude is investigated through parameter variation. The selling of surplus electricity from the contracted renewable PPAs is not implemented in the optimisation but is subject to complementary results analyses and discussions. In the course of this study, analyses are performed in two European bidding zones: in the German bidding zone (DE) and in the Spanish bidding zone (ES). Since there are currently no notable electrical grid fees or taxes for electrolyser operation in Germany, they are assumed to be zero [23], [24]. However, a currently ongoing revision of the grid fee concept could lead to substantial fees in the near future. In Spain, grid fees and non-recoverable taxes apply for large-scale electrolyser operators (above 150,000 MWh/a electricity consumption) of around 7.6 €/MWh to 8.1 €/MWh [25]. Applying those grid fees and taxes to the analysis in Spain would close the PPA price gap between Spain, with higher onshore wind and photovoltaic yields, and Germany, even though it could widen in the near future once the German grid fee concept is revised. Since the price gap is an essential difference between European countries, the grid fees and taxes in Spain are also assumed zero to ensure a “level-playing-field” between the available PPA options across the different bidding zones. For electrolyser and compressor sizing, annualised installation cost incur. The considered hydrogen storage options are allocating a salt cavern storage bundle and installing a pressure gas tank.

Where not differently pointed out, the time characteristic of the predefined hydrogen demand is chosen to be flat, mirroring the demand of large-scale industrial customers, which are the frontrunners in using locally produced green hydrogen in Europe [16], [26]. The annual hydrogen demand to be met is 36,500 tonnes per year, which corresponds to the size of currently launched green hydrogen production projects of approximately 300 MW electrolysers, providing for large-scale industrial consumers [26]. Supplementary Note 1 in the supplementary data provides a list of all economic and technical parameters. For this study, all economic input parameters are set to present cost values derived from recent publications and converted into euros (2024 value, $€_{2024}$) using the Chemical Engineering Plant Cost Index (CEPCI) [27]. The formulated optimisation problem, including the objective function and constraints, is linear, and the optimisation time frame is one year at hourly resolution. Section 6 in the Appendix includes the complete mathematical formulation.

## 2.2 Design paradigms

To show the effects of the additionality criterion and associated long-term renewable PPA contracts on green hydrogen production costs and the fulfilment of offtake agreements, three different design paradigms are introduced (see Figure 2).

Applying the reference design paradigm (see Figure 2 (a)) to the hydrogen production system, begins with selecting a reference year with medium renewable power availability from the past 25 weather years. The selection is based on capacity factor time series for all PPA options. Hereby, the full set of time series for the past 25 years captures the inter-year uncertainty in power feed-in associated with long-term renewable PPA contracts. Supplementary Note 1 in the supplementary data contains details on the reference year selection. The time series for the reference year are used as an input to the hydrogen production system optimisation introduced in Section 2.1 ((see Figure 2 (a) step 1.). The resulting system design from this optimisation contains the cost-optimal sizing of all system components, henceforth referred to as the reference design. Subsequently, as depicted in Figure 2 (a) step 2., the reference design is used as a precondition for the hydrogen production system optimisation, which is repeated 24 times, each time using one of the other 24 feed-in time series from the full set as an input. The results enable an impact evaluation of the feed-in uncertainties associated with long-term PPA contracts.

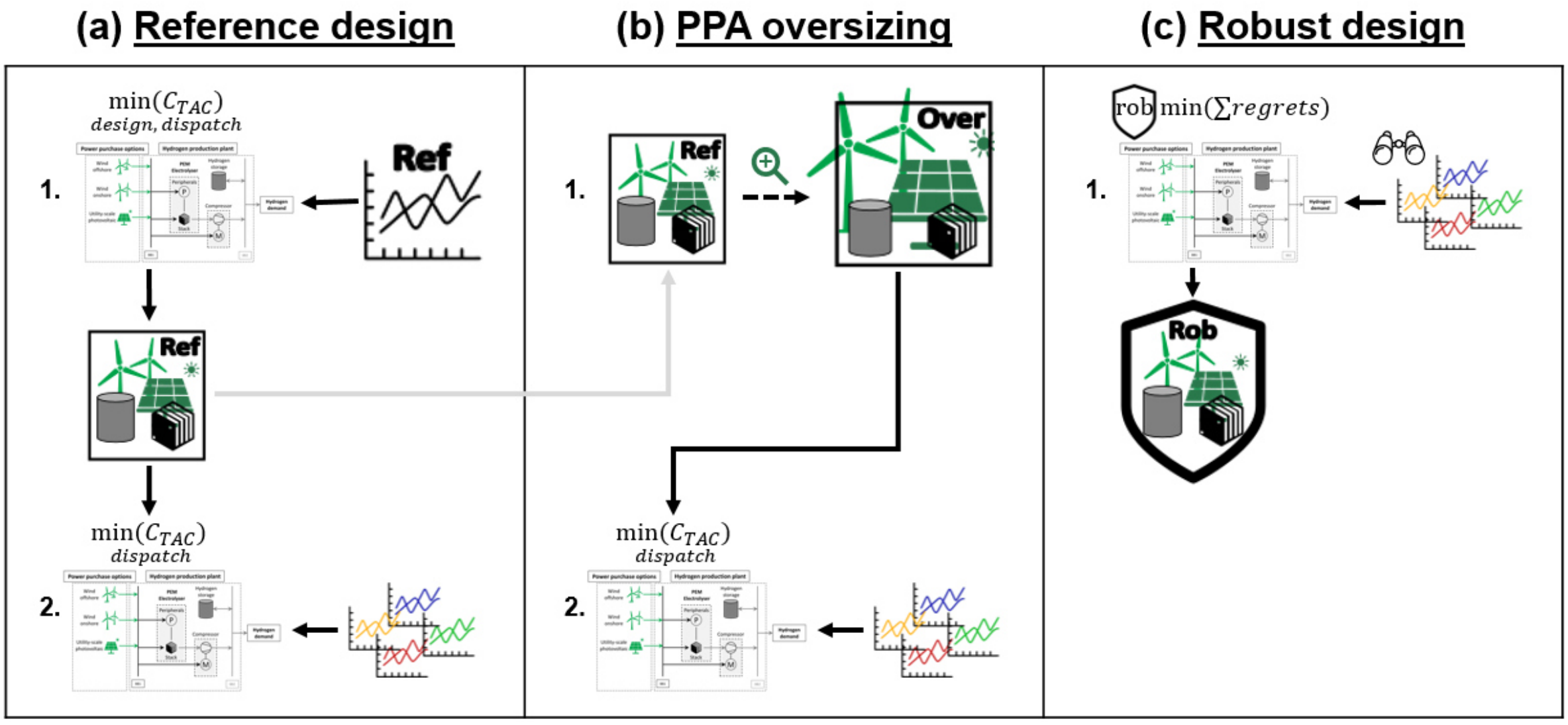

*Figure 2: Analysed design paradigms of green hydrogen production systems. (a), Reference design paradigm. (b), PPA oversizing design paradigm. (c), Robust design paradigm.*

Applying the PPA oversizing paradigm (see Figure 2 (b)) and the robust design paradigm (see Figure 2 (c)) enables an evaluation of different electrolyser operator design measures to mitigate the volume risk associated with additional renewable PPAs.

An apparent measure to address potential volume issues arising from inter-year feed-in fluctuations is to oversize PPA capacities. Therefore, the reference PPA design is increased by 10 % and 20 % to compensate for missing energy in underproduction years (see Figure 2 (b) step 1.). Subsequently, the reference design with oversized PPA capacities is used as a precondition for the hydrogen production system optimisation, which is run for all 25 feed-in years (see Figure 2 (b) step 2.). Section 6.1.2 includes details on PPA oversizing.

To provide a benchmark for evaluating the design choices under the reference design and PPA oversizing paradigms, a robust design optimisation approach is used. The chosen optimisation method extends the hydrogen production system optimisation by perfect foresight over all 25 individually considered weather years (see Figure 2 (c) step 1.). The resulting system design ensures offtake fulfilment in all 25 years at minimal cost. The chosen method for robust optimisation is oriented on the concept of "optimality robustness" [28]. Different established robust optimisation approaches, for example the "Worst-case optimisation" [29], were also tested and provided similar results. Details on the chosen method and mathematical details are in Section 6.1.3.

## 2.3 System modifications

Since the results of techno-economic assessments in similar studies (see Section 0) depend on techno-economic assumptions, the five different system modifications in Table 1 are applied to the considered hydrogen production system. Combined with the evaluation of the previously introduced design paradigms, they provide a comprehensive picture of the effects of the additionality criterion and associated long-term contracts on green hydrogen production and the effectiveness of different measures for volume risk mitigation (see Table 1 – Risk mitigation measures analysed). Each Section 3.1 to 3.5 includes the results of one system modification (see Table 1 – Result Section)

*Table 1: Overview of result Sections, made system modifications and analysed risk mitigation measures*

| Results Section | System modification | Risk mitigation measures analysed |
|---|---|---|
| 3.1 | Standard system configuration (no modification) | Operator design measures |
| 3.2 | Increased storage cost | Operator design measures |
| 3.3 | Different bidding zone | Operator design measures |
| 3.4 | Relaxed rules for temporal correlation | Operator design and regulatory measures |
| 3.5 | Increased offtake flexibility | Operator design and offtaker measures |

First, in Section 3.1, all design paradigms are evaluated for a standard system configuration of the hydrogen production system to create a basic understanding of the effects of the additionality criterion and the effectiveness of different operator design measures to mitigate potential volume risks. The central assumptions of the standard configuration are the salt cavern storage bundle, the German bidding zone (DE), hourly temporal correlation and a flat hydrogen offtake. The subsequent modifications are all based on this standard configuration.

To broaden the picture of possible effects, Sections 3.2 and 3.3 include results for two system modifications: increased storage costs and a different bidding zone in Europe. Since storage costs are an impactful economic factor in green hydrogen production when matching a fluctuating renewable power source to an offtake characteristic [30] and salt caverns are not yet widely available for all production projects due to infrastructural limitations, a more expensive storage option is analysed. Due to the geographical correlation criterion, the location of an electrolyser affects its power sourcing options because the availability and quality of renewable PPA options differ across Europe's bidding zones. Therefore, in addition to DE, the Spanish bidding zone (ES) is analysed. In ES, different PPA options with different characteristics compared to DE are available, large green hydrogen projects are currently being implemented, with future national expansion ambitions in green hydrogen production [31], [32], and comparable hydrogen storage options are available as in Germany [33], [34], making ES a suitable comparator to DE.

Since the rules for temporal correlation are subject to controversial discussions, and to provide a better understanding of the interactions between the different power sourcing criteria, the rules for temporal correlation are relaxed from hourly to monthly correlation in Section 3.4. This relaxation of the temporal correlation condition currently applies for first mover projects in a transitional phase until 2030. The extension of this transitional rule is part of the proposed adjustments in the joint letter of ministries from European Union member states calling for a preliminary review of the power sourcing criteria.

Finally, in Section 3.5, the hydrogen offtake characteristics are modified to emulate a more flexible green hydrogen demand, which can be advantageous regarding the mitigation of volume risks posed by long-term renewable PPA contracts. This is especially relevant in the upcoming years, where blending green hydrogen is possible in certain applications.

# 3 Results and Discussion

## 3.1 Standard system configuration

To establish a basic understanding of the effects of the additionality criterion and the effectiveness of different operator design measures to mitigate potential volume risks, the design paradigms presented in Section 2.2 are applied to the standard configuration (see Section 2.3) of the hydrogen production system introduced in Section 2.1.

Figure 3 (a) depicts the share of years with unfulfilled offtake agreement on the left y-axis (yellow star) and the share of total demand needed to meet the offtake agreements in years with unfulfilled offtake agreement on the right y-axis (blue violin) under all analysed design paradigms. Figure 3 (b) shows the resulting *OHSC* as violin plots, and Figure 3 (c) shows the nominal sizes of system components for all design paradigms.

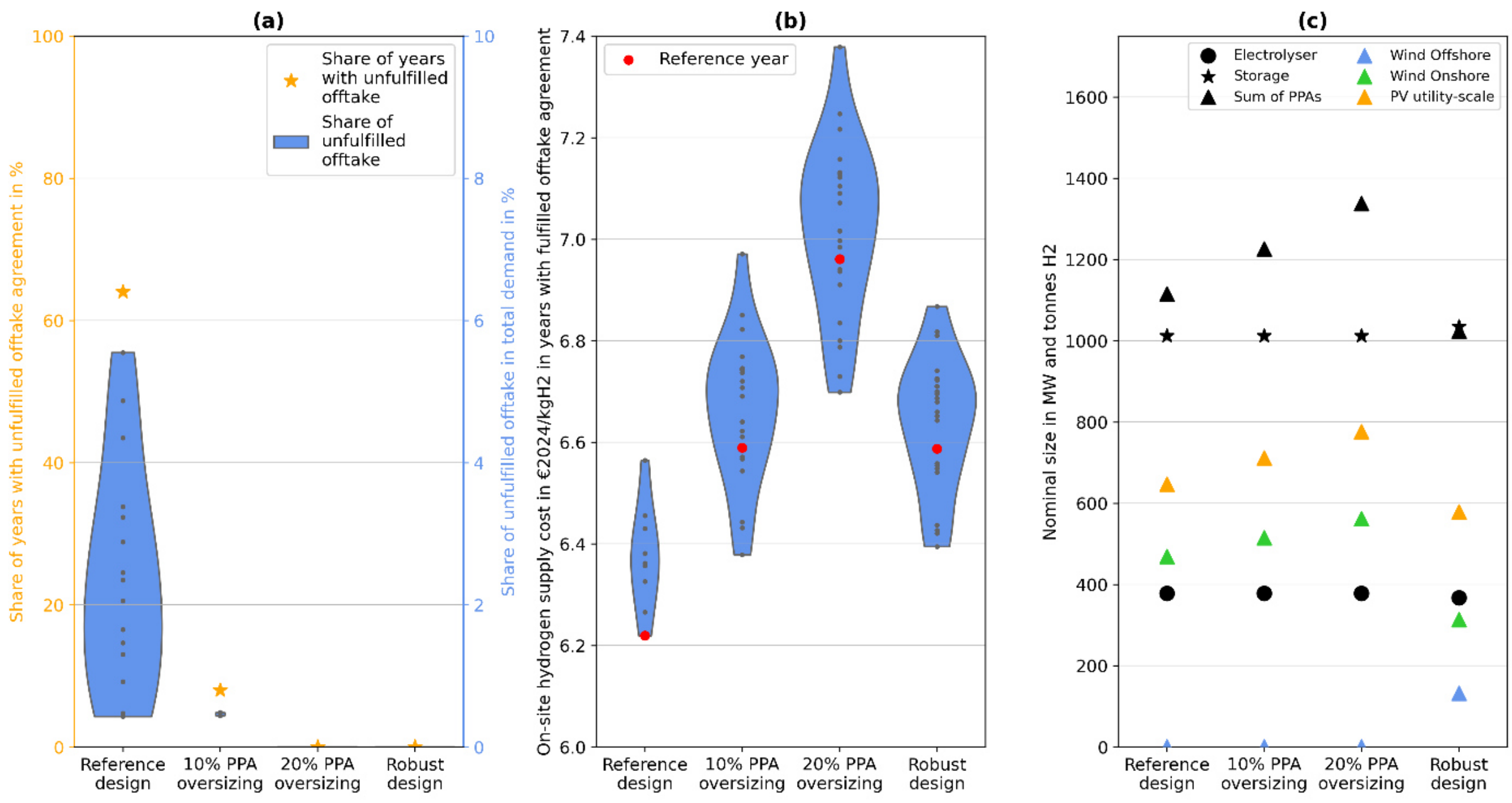

*Figure 3: Optimisation results for the standard configuration of the considered green hydrogen production system under all analysed design paradigms. (a), yellow y-axis: Share of years with unfulfilled offtake agreement out of all analysed years; blue y-axis: Share of total demand needed to meet unfulfilled offtake agreement in years with unfulfilled offtake agreement. (b), On-site hydrogen supply cost in years with fulfilled offtake agreement. (c), Nominal*

*sizes of electrolyser, storage and PPAs. The storage option is a salt cavern storage bundle. The bidding zone is Germany (DE). The temporal correlation is hourly. The hydrogen offtake characteristic is flat.*

Figure 3 (a) shows that, in the reference design optimisation, in 64 % of the 25 analysed years, the predefined hydrogen demand cannot be met, with up to 5.5 % of the offtake unfulfilled due to inter-year fluctuations in renewable PPA power availability. However, oversizing the PPA options of the reference design by 10 % reduces the share of years with unfulfilled offtake to 8 %, and leaving only marginal amounts of unfulfilled offtake. Although oversizing the PPA options by 20 % resolves the issue of unfulfilled offtake agreements, the PPA overbooking has notable effects on the *OHSC*, as shown in Figure 3 (b). Each 10 % PPA upsizing step results in an *OHSC* increase of 0.37 €/kgH2 in the reference year. The resulting spread of *OHSC*, caused by inter-year fluctuations in PPA power availability, ranges between 0.35 €/kgH2 under the reference design paradigm and 0.68 €/kgH2 in the 20 % PPA oversizing paradigm for all years with a fulfilled offtake agreement. Even though the cost-optimal robust system design leads to *OHSC* similar to 10 % PPA oversizing (see Figure 3 (b)), the robust system design in Figure 3 (c) is different. In contrast to a simple PPA upsizing, the cost-optimal robust design is achieved by PPA diversification combined with slight electrolyser downsizing and storage up-sizing. Although not included in the reference design due to its high price, the offshore wind option is chosen to increase the system's robustness against inter-year feed-in fluctuations because of its particularly high capacity factor (for details see Supplementary Note 1 in the supplementary data).

Since PPA oversizing may affect the surplus energy production of the PPAs, and the performed optimisations do not include surplus selling to the electricity market, which in turn can affect the *OHSC*, this aspect is explored further. Figure 4 (a) depicts the share of annual surplus in total PPA production under all design paradigms in the years with fulfilled offtake agreement. Based on the *OHSC* in the years with fulfilled offtake agreement in Figure 4 (b) (same as Figure 3 (b)), Figure 4 (c) shows the corrected *OHSC* after surplus selling as a function of the annual average selling price. Here the thick lines in the coloured areas indicate the reference year and the thin boundary lines the respective maximum and minimum values. Calculation details are in Section 6.3 in the Appendix, equations (39), (40) and (41).

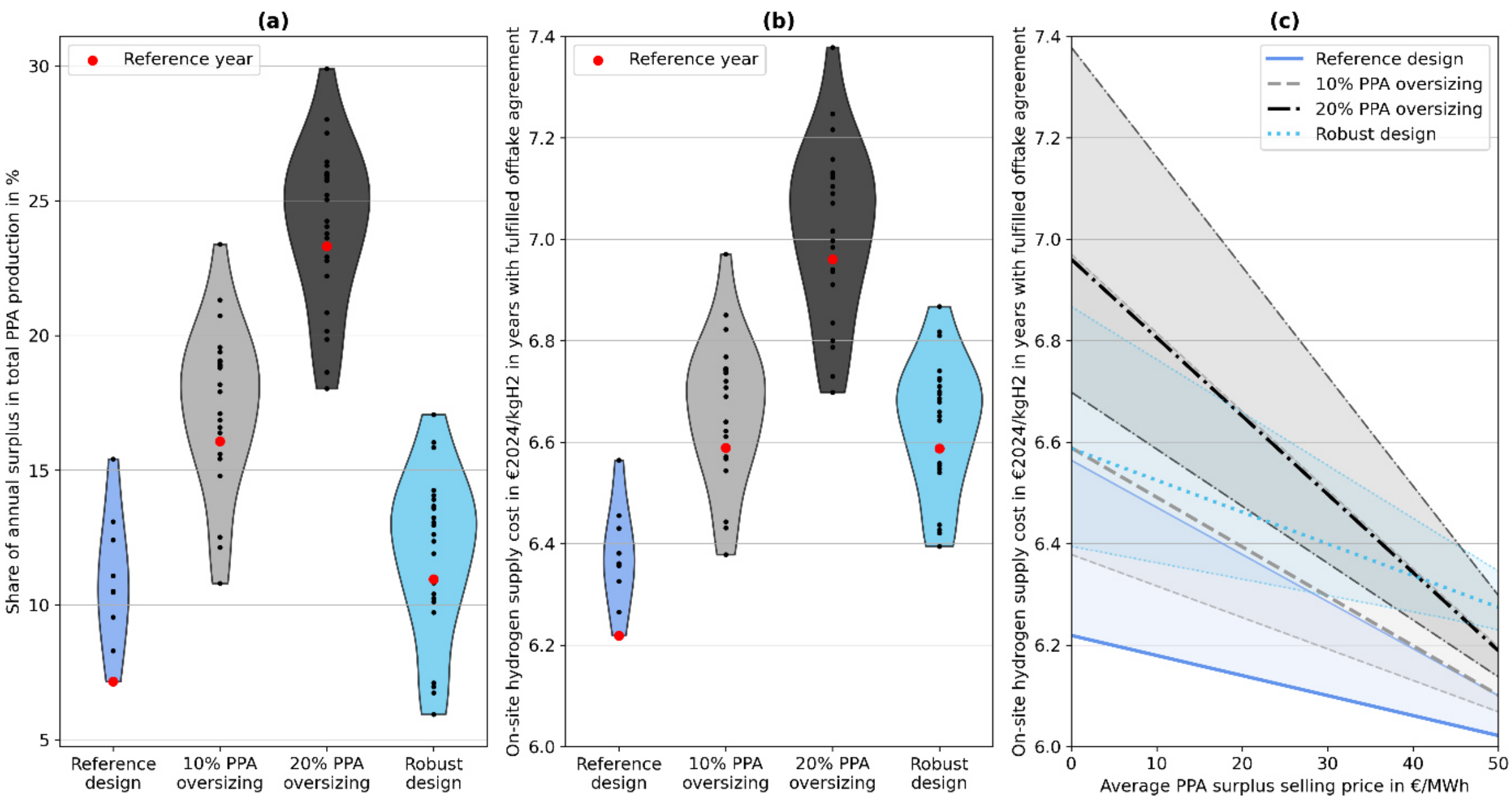

*Figure 4: Optimisation results for the standard configuration of the considered green hydrogen production system under all analysed design paradigms with additional PPA surplus selling. (a), Share of annual surplus in total PPA production in years with fulfilled offtake agreement. (b), On-site hydrogen supply cost in years with fulfilled offtake*

*agreement. (c), On-site hydrogen supply cost in years with fulfilled offtake agreement corrected by PPA surplus selling revenues as a function of the PPA surplus selling price. Thick lines within coloured areas in (c) show reference-year values, thin lines on area boundaries show minimum and maximum values. The storage option is a salt cavern storage bundle. The bidding zone is Germany (DE). The temporal correlation is hourly. The hydrogen offtake characteristic is flat.*

As expected, increasing PPA sizes leads to a notable increase in surplus PPA energy (see Figure 4 (a)). Considering the reference year, the surplus energy more than doubles for 10 % PPA upsizing and more than triples for 20 % PPA upsizing compared to the reference design. Despite the integration of the offshore wind option with comparably high capacity factors and the resulting increase in *OHSC* (see Figure 4 (b)), the surplus amounts in the robust design are only slightly higher than in the reference design. As shown in Figure 4 (c), selling surplus energy can notably reduce *OHSC* in general, with the magnitude of the reduction directly related to the average selling price. Due to the high amounts of PPA surplus under the PPA oversizing paradigms, the magnitude of possible *OHSC* reductions is comparably high. It can partly compensate for the cost increase from PPA overbooking. Comparing selling prices of 0 €/MWh and 50 €/MWh and considering the reference year, the *OHSC* gap between the reference design (solid dark-blue line, thick) and 20 % PPA oversizing (dash-dotted black line, thick) can be reduced from 0.74 €/kgH2 to 0.17 €/kgH2. However, assuming the majority of renewable energy sources in the same bidding zone mirror times with high feed-in of the booked renewable PPAs, increased annual average selling prices for renewable surplus seem unlikely.

In summary, considering the standard configuration of the analysed green hydrogen production system, inter-year fluctuations of renewable PPAs feed-in volumes notably affect on-site hydrogen supply cost and can jeopardise the fulfilment of green hydrogen offtake agreements. A simple overbooking of renewable PPA capacities provides a suitable method to address the feed-in volume issue, but notably increases supply cost by increasing surplus PPA production. In contrast to a simple PPA overbooking, PPA diversification and storage upsizing are the cost-optimal electrolyser operator measures to increase the robustness of green hydrogen production systems.

## 3.2 Increased storage costs

To explore the effects of increased storage costs, which have proven to be an impactful economic driver of green hydrogen production [30], the salt cavern storage bundle is replaced with a pressure tank. Figure 5 (a1), (b1) and (c1) show the same as Figure 3 (a), (b) and (c) as a reference, Figure 5 (a2), (b2) and (c2) show the respective counterparts for increased storage cost.

Comparing Figure 5 (a1) and (a2) shows an increase in the share of years with unfulfilled offtake agreements across all design paradigms. Under the reference design paradigm, the high storage costs lead to a 12 % increase, and the 20 % PPA upsizing is no longer sufficient to solve the issue of unfulfilled offtake agreements. In contrast, the share of unfulfilled offtake in the total demand falls below 1 % across all design paradigms. As one might expect, the increase in storage costs results in higher *OHSC* (see Figure 5 (b1) and (b2)). They increase by around 1.25 €/kgH2 across all design paradigms in the reference year. The additional cost for 20 % PPA upsizing results in 0.96 €/kgH2 *OHSC* increase. Examining the component sizing in Figure 5 (c1) and (c2) under the reference design paradigm shows that increased storage costs incentivise notable storage downsizing, combined with offshore wind integration. The small storage capacity limits the ability to buffer intra-year fluctuations in power availability, leading to more years with unfulfilled offtake agreement in Figure 5 (a2). Counteracting this is the integration of expensive but reliable offshore wind power, whose effect, however, becomes apparent in the notably decreased share of unfulfilled offtake in Figure 5 (a2). To achieve cost-optimal robustness, the storage size is increased from 267 tonnes H2 to 450 tonnes H2, allowing the PPA portfolio composition to shift towards the cheaper onshore wind and photovoltaic options (see Figure 5 (c2)).

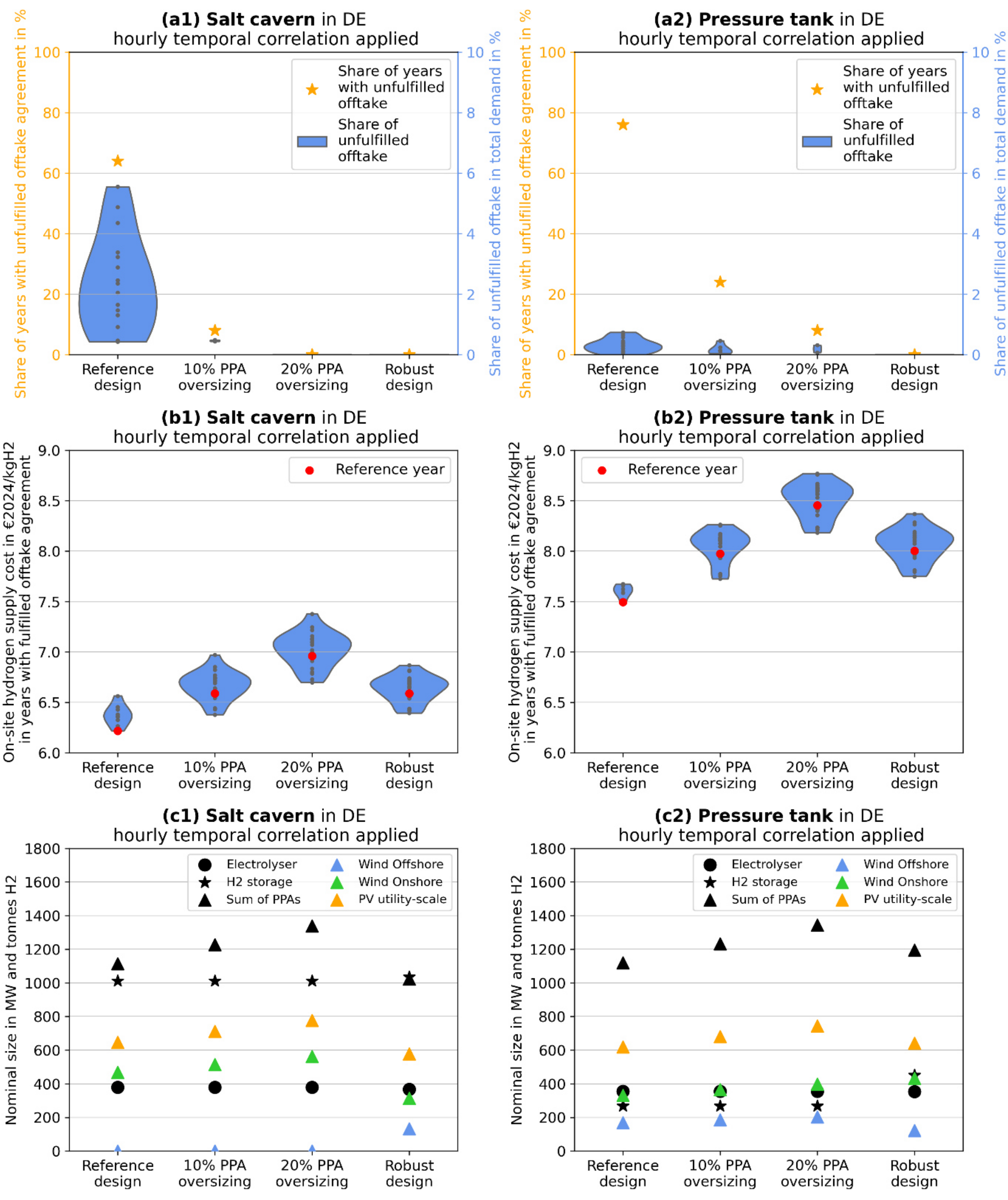

*Figure 5: Comparison of optimisation results between the standard configuration of the considered green hydrogen production system (a1-c1) and increased storage cost (a2-c2) for all analysed design paradigms. (a1) and (a2), yellow y-axis: Share of years with unfulfilled offtake agreement out of all analysed years; blue y-axis: Share of total demand needed to meet unfulfilled offtake agreement in years with unfulfilled offtake agreement. (b1) and (b2), On-site hydrogen supply cost in years with fulfilled offtake agreement. (c1) and (c2), Nominal sizes of electrolyser, storage and PPAs. The storage option in the standard configuration (a1-c1) is a salt cavern storage bundle, for increased storage cost (a2-c2) it is a pressure gas tank. The bidding zone is Germany (DE). The temporal correlation is hourly. The hydrogen offtake characteristic is flat.*

The key findings on increased storage costs confirm those previously made regarding the effects of the additionality criterion and associated inter-year PPA power feed-in fluctuations. In contrast to lower storage costs, electrolyser operator measures to achieve cost-optimal robustness should primarily focus on storage upsizing, which, in turn, enables prioritising cheap PPA options in a diverse portfolio. However, the choice of increased storage cost options, such as pressure gas tanks, for large-scale electrolyser projects remains questionable due to the notably increased *OHSC* and possible technical limitations, which in turn emphasises the importance of widely accessible, low-cost hydrogen storage options.

### 3.3 Different bidding zone

Given the differences in PPA availability and characteristics, the Spanish bidding zone (ES) is selected to extend the applicability of findings across Europe. In ES, there are currently no offshore wind options available, whereas the onshore wind and photovoltaic options have notably higher expected yields than those in DE. Figure 6 shows the same as Figure 5, expect that Figure 6(a2), (b2), and (c2) show the results for the Spanish bidding zone.

As similarly explored for increased storage costs (see Figure 5 (a2)), the change in bidding zone increases the share of years with unfulfilled offtake agreements relative to the standard configuration (see Figure 6 (a1) and (a2)). In this case, from 64 % to 84 % in the reference design. Contrastingly, the resulting share of unfulfilled offtake only slightly decreases under the reference design paradigm but increases to above 1 % for both PPA oversizing paradigms. The lower prices and higher yields of PPA options, especially for photovoltaic, reduce the *OHSC* for the reference design and PPA oversizing paradigms by between 0.31 €/kgH2 (reference design) and 0.36 €/kgH2 (20 % PPA oversizing) compared to the standard configuration in the reference year (see Figure 6 (b1) and (b2)). Under the robust design paradigm, on the other hand, the reduction is only 0.11 €/kgH2. Comparing the PPA and storage sizing in Figure 6 (c1) and (c2) under the reference design paradigm shows notable downsizing across all technologies. Especially noticeable here is a shift in the PPA portfolio towards a more photovoltaic-based portfolio, driven by the increased price gap relative to the onshore wind option (for details see Supplementary Note 1 in the supplementary data). This combination of a more photovoltaic-based PPA portfolio and reduced storage capacity notably decreases the flexibility for intra-year energy shifting when inter-year differences arise, thereby diminishing the robustness of the production system, as seen in Figure 6 (a1). Since no additional power purchase option is available for PPA portfolio diversification, a twofold increase in storage capacity, and a slight adjustment in the PPA portfolio composition in favour of the onshore wind option lead to cost-optimal system robustness. The large increase in storage capacity needed to achieve robustness relative to DE, explains the comparatively small *OHSC* reduction under the robust design paradigm (see Figure 6 (b1) and (b2)).

In summary, despite positive effects on *OHSC*, the availability of a cheap, high-yield photovoltaic option rather than a more expensive but also more diverse PPA portfolio exacerbates the issue of unfulfilled offtake agreements caused by long-term renewable PPA contracts. Furthermore, it requires above-average storage capacities to ensure system robustness against inter-year feed-in fluctuations. This, in turn, can offset the advantages of a bidding zone with high-yield renewable energy sources. However, a portfolio diversification through the combination of different onshore wind or photovoltaic locations could help to solve this issue.

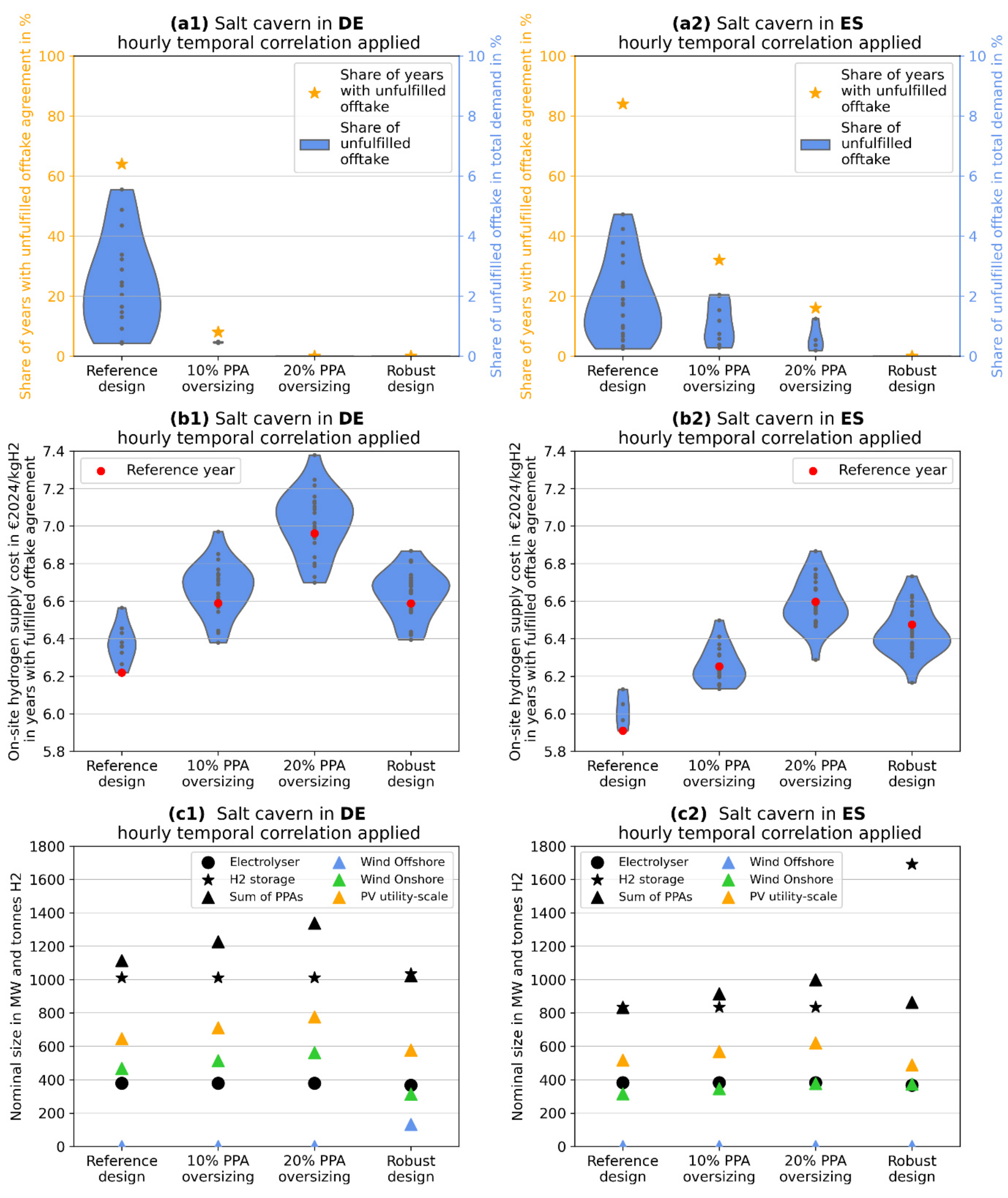

*Figure 6: Comparison of optimisation results between the standard configuration of the considered green hydrogen production system in the German (DE) bidding zone (a1-c1) and in the Spanish (ES) bidding zone (a2-c2) for all analysed design paradigms. (a1) and (a2), yellow y-axis: Share of years with unfulfilled offtake agreement out of all analysed years; blue y-axis: Share of total demand needed to meet unfulfilled offtake agreement in years with unfulfilled offtake agreement. (b1) and (b2), On-site hydrogen supply cost in years with fulfilled offtake agreement. (c1) and (c2), Nominal sizes of electrolyser, storage and PPAs. The storage option in both bidding zones is a salt cavern storage bundle. The temporal correlation is hourly. The hydrogen offtake characteristic is flat.*

## 3.4 Relaxed rules for temporal correlation

With the preliminary review of the power purchase criteria for green hydrogen production in 2026, understanding their interactions is crucial for developing a comprehensive picture of their effects on the economics and emissions of green hydrogen production. To demonstrate how a switch from hourly to monthly temporal correlation affects the explored issue of unfulfilled offtake agreements arising from the additionality criterion, the option of limited grid electricity integration is added to the standard configuration of the green hydrogen production system. The additional mathematical constraints for monthly temporal correlation are given in equation (37).

Figure 7Figure 6 shows the same as Figure 5, except that Figure 6(a2), (b2), and (c2) show the results for monthly instead of hourly temporal correlation. The annual average grid electricity price for additional grid electricity mix integration is the average electricity price in Germany for non-household consumers in 2024 with a consumption of 150 GWh or above [25].

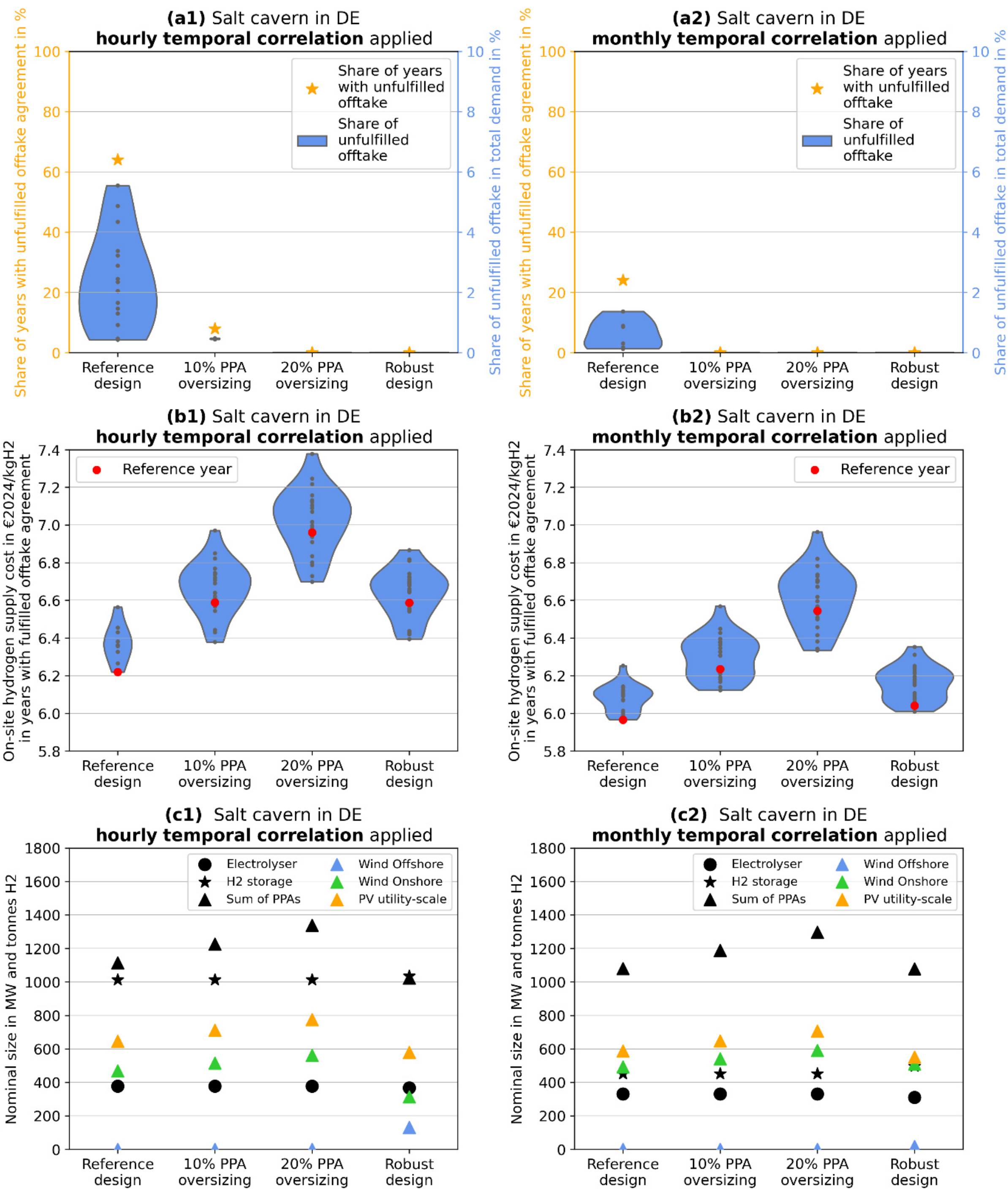

*Figure 7: Comparison of optimisation results between the standard configuration of the considered green hydrogen production system with hourly (a1-c1) and monthly temporal correlation (a2-c2) for all analysed design paradigms. (a1) and (a2), yellow y-axis: Share of years with unfulfilled offtake agreement out of all analysed years; blue y-axis: Share of total demand needed to meet unfulfilled offtake agreement in years with unfulfilled offtake agreement. (b1)*

*and (b2), On-site hydrogen supply cost in years with fulfilled offtake agreement. (c1) and (c2), Nominal sizes of electrolyser, storage and PPAs. The storage option is a salt cavern storage bundle. The bidding zone is the German (DE) bidding zone. The hydrogen offtake characteristic is flat. The average electricity price for non-household consumers in 2024 in Germany with a consumption of 150 GWh or above is used as average grid electricity price for additional grid electricity purchase. [27]*

Comparing Figure 7 (a1) and (a2) shows that adjusting temporal correlation from hourly to monthly notably reduces the share of years with unfulfilled offtake agreements and the share of unfulfilled offtake in total demand. Considering the reference design paradigm, the share of years with unfulfilled offtake agreements decreases from 64 % to 24 %, and the share of unfulfilled offtake is below 1.75 % in all those years. At 10 % PPA oversizing, no years remain with unfulfilled offtake agreements. The resulting *OHSC* in Figure 7 (b2) are between 0.25 €/kgH2 in the reference design and 0.55 €/kgH2 in the robust design below the *OHSC* in the standard configuration in Figure 7 (b1) in the reference year. In contrast to increased storage costs and changed bidding zone (see Sections 3.2 and 3.3), the *OHSC* in the cost-optimal robust design shifts more towards the *OHSC* magnitude of the reference design and its spread is reduced from 0.47 €/kgH2 to 0.34 €/kgH2. The system sizing in Figure 7 (c1) and (c2) shows that integrating a flexible power source enables reduction in electrolyser size and a more than halving of storage sizing across all design paradigms. To achieve cost-optimal robustness under the robust design paradigm, the system design modifications compared to the reference design in Figure 7 (c2) are marginal. Slightly decreasing the electrolyser and increasing the storage sizing combined with a minimal shift of the PPA portfolio towards the wind options, including minimal offshore wind volumes, does the trick.

Since the economic incentive for grid electricity integration and the respective effects on system design depend on its price, the annual average grid electricity price is varied between 40 €/MWh and 180 €/MWh, and the optimisation is repeated for all design paradigms. Figure 8 (a) to (f) shows the price variation on the x-axis and the previously assumed German average price as a reference. Figure 8 (a) depicts the share of years with unfulfilled offtake agreement (yellow stars in Figure 7 (a2)), and (b) depicts the share of unfulfilled offtake agreement (blue violin plots in Figure 7 (a2)). In (b), the thick lines in the coloured areas indicate the mean, and the thin boundary lines the respective maximum and minimum values. As the primary motivation of the power purchase rules is to avoid increased emissions, and notable emissions are assigned to the grid electricity mix in most European countries [35], Figure 8 (c) shows the resulting emissions across all design paradigms in years with fulfilled offtake agreements. Here the thick lines in the coloured areas indicate the reference year and the thin boundary lines indicate the respective maximum and minimum values. The minimum saving threshold for green hydrogen of 70 % over grey hydrogen is depicted as an additional reference [36]. To better understand the changes in system robustness in the reference design in (a) and (b) and the effects on the resulting emissions, (d) shows the component sizing of the production system, (e) shows the annual sum of purchased energy from the different power purchase options and (f) shows the utilisation of the electrolyser in annual full load hours. All for the reference design paradigm. In (e) and (f), the thick lines in the coloured areas indicate the reference-year value and the thin boundary lines the respective maximum and minimum values. Details on the emission intensity and electrolyser utilisation calculations are in Section 6.3 equations (42) and (43).

Focusing on reduced grid electricity prices relative to the German average, an increase in years with unfulfilled offtake and an increase in the share of unfulfilled offtake are depicted in Figure 8 (a) and (b), especially under the reference design paradigm. The resulting numbers are between the results for hourly and monthly temporal correlation in Figure 7 (a1) and (a2). Figure 8 (c) shows that reduced prices for grid electricity incentivise a slightly lower PPA sizing, with a portfolio composition shifting towards photovoltaic integration. Consequently, the annual amount of energy available from all PPAs decreases (see Figure 8 (e)). To fulfil the offtake agreement in the reference year, increased storage capacity (see Figure 8 (d)) and increased grid electricity integration (see Figure 8 (e)) are needed. This, in turn, enables downsizing of the electrolyser (see Figure 8 (d)) and results in an increase in electrolyser utilisation (see Figure 8 (f)). The reduced electrolyser sizing counteracts the system flexibility for intra-year energy shifting gained from storage upsizing. Combined with the lowered amounts of available PPA power, which limit the flexibility for monthly grid electricity mix integration through the nature of the temporal correlation criterion, this impairs the robustness of the production system to compensate inter-year differences in PPA power feed-in, which is shown in Figure 8 (a) and (b). The increased integration of grid electricity mix at lower price levels results in an increase in emission intensity across all design paradigms, with the highest emissions assigned to the reference design (see Figure 8 (c)). The minimum saving threshold for green hydrogen is only exceeded by a single year in the reference design at 40 €/MWh.

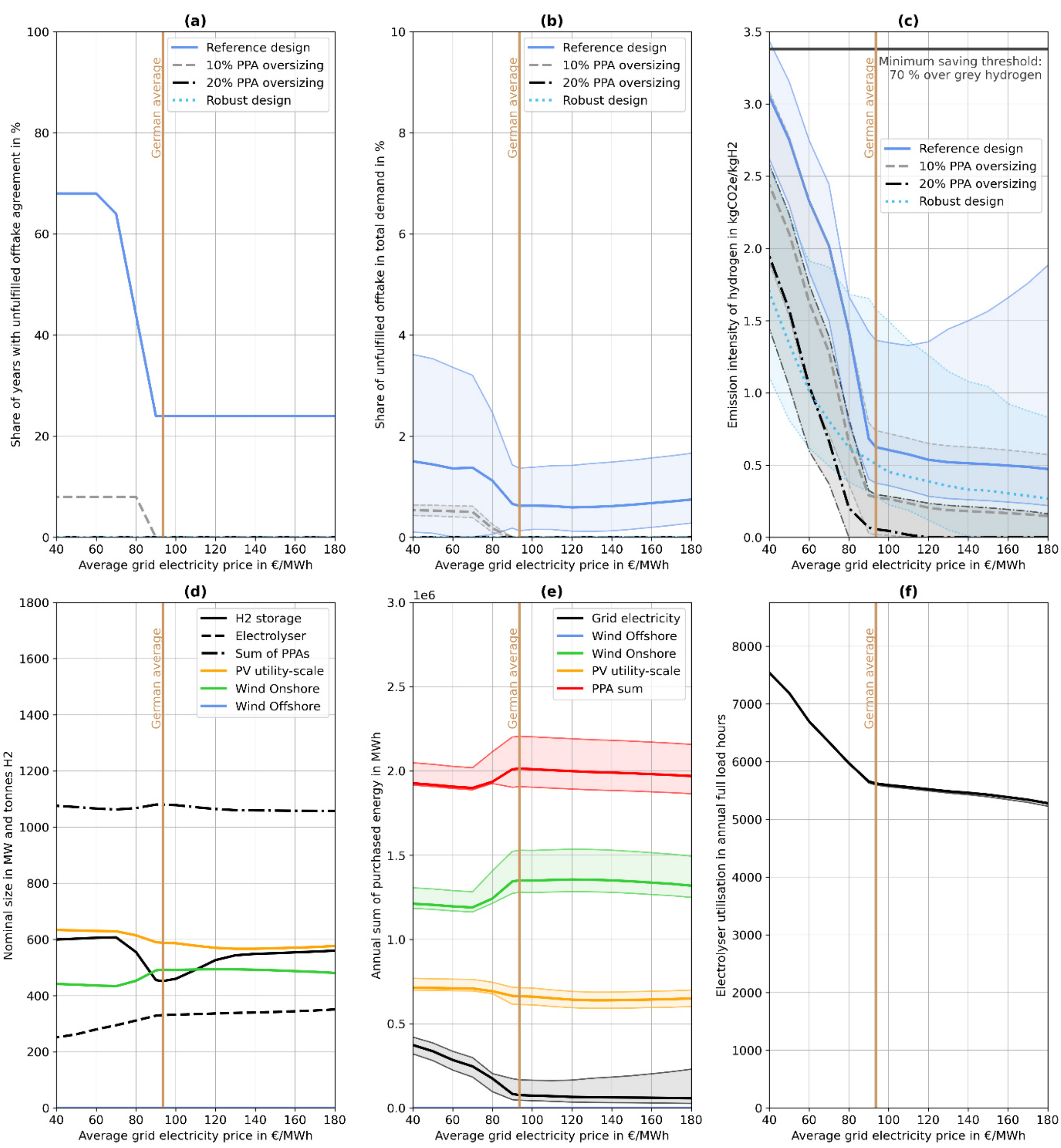

*Figure 8: Optimisation results for the modified green hydrogen production system with monthly instead of hourly temporal correlation. (a), Share of years with unfulfilled offtake agreement out of all analysed years as a function of the average grid electricity price. (b), Share of total demand needed to meet unfulfilled offtake agreement in years with unfulfilled offtake agreement as a function of the average grid electricity price. (c), Emission intensity of the*

*produced hydrogen as a function of the average grid electricity price in years with a fulfilled offtake agreement. (d), Nominal sizes of electrolyser, storage and PPAs under the reference design paradigm as a function of the average grid electricity price. (d), Annual sum of purchased power from the different power purchase options in the reference design paradigm as a function of the average grid electricity price. (e), Utilisation of the electrolyser in full load hours under the reference design paradigm as a function of the average grid electricity price. Thick lines within coloured areas in (b) show mean, thin lines on area boundaries show maximum and minimum values. Thick lines within coloured areas in (c), (e) and (f) show reference-year values, thin lines on area boundaries show minimum and maximum values. The storage option is a salt cavern storage bundle. The bidding zone is Germany (DE). The hydrogen offtake characteristic is flat. (a)-(f) additionally show the average electricity price in Germany for non-household consumers in 2024 with a consumption of 150 GWh or above as a reference* [25]. *In addition, (c) displays the minimum emission intensity saving threshold set for green hydrogen, which is 70 % over its fossil comparator grey hydrogen* [37].

Increasing grid electricity prices relative to the German average does not affect the share of years with unfulfilled offtake agreements across all design paradigms (see Figure 8 (a)). The share of unfulfilled offtake is only slightly increasing under the reference design paradigm shown in Figure 8 (b). The system design changes in Figure 8 (d) show that the higher the prices of grid electricity mix, the more favourable a combination of increased electrolyser sizing and increased storage capacity is for structuring fluctuating PPA power availability to fulfil the flat offtake, rather than using expensive grid electricity. Higher storage capacities, in turn, enable slightly greater integration of cheap photovoltaic power, which, on the downside, slightly increases the shares of unfulfilled offtake in Figure 8 (b) and necessitates greater grid electricity mix integration in years with PPA underproduction. This causes increased emissions in some years under the reference design paradigm (see Figure 8 (c), maximum values, solid blue line). In contrast, the emission intensity is generally decreasing with increasing prices across all design paradigms.

In conclusion, relaxing the rules on temporal correlation is a suitable regulatory measure to increase the robustness of green hydrogen production systems by mitigating the volume risk posed by the additionality criterion and resulting long-term PPA contracts. By avoiding costly PPA and storage upsizing, and PPA portfolio diversification, notable *OHSC* reductions can be achieved, and the *OHSC* spread caused by inter-year fluctuations can be narrowed. However, the level of grid electricity prices decisively affects the effectiveness of relaxed temporal correlation rules. At low prices, the nature of the monthly temporal correlation criterion, which ties potential additional power purchase to the inter-year fluctuating PPA yield, which again is caused by the additionality criterion and long-term contracts, limits the flexibility in additional power purchase and thus partly offsets the positive effects on system robustness. Furthermore, enabling grid electricity mix integration through relaxed temporal correlation rules results in production-related emissions below the minimum saving threshold relative to grey hydrogen, except in single years with annual average grid electricity prices below 50 €/MWh.

To provide a comprehensive picture of the interactions between the different power sourcing criteria, the computations are repeated for the Spanish bidding zone (ES). The results generally confirm the findings for DE, whereas the effects of relaxed temporal correlation on system robustness are less and on *OHSC* are more impactful at average grid electricity prices. The resulting emission intensity is substantially lower across all design paradigms and electricity prices, mainly due to the lower average emission intensity of grid electricity in ES. The detailed results are in Supplementary Note 3 in the supplementary data.

## 3.5 Increased offtake flexibility

Assuming a flat offtake characteristic is based on the largest future hydrogen applications in Europe, such as steel producers, and existing ones, which are refineries, the ammonia industry, and other chemical processes [16], [38]. Here, grey hydrogen from steam methane reforming is currently used but shall be replaced by green hydrogen to achieve the European Union's climate target goals [36]. However, some applications may provide more flexibility than covered by a flat green hydrogen offtake profile. This could occur in the upcoming years, when green hydrogen accounts for only a fraction of a process's total hydrogen demand and flexibilities in grey hydrogen provision exist. Furthermore, other potential future applications, such as industrial heat provision, power plant operation and different players from the transport sector may offer greater flexibility in their offtake.

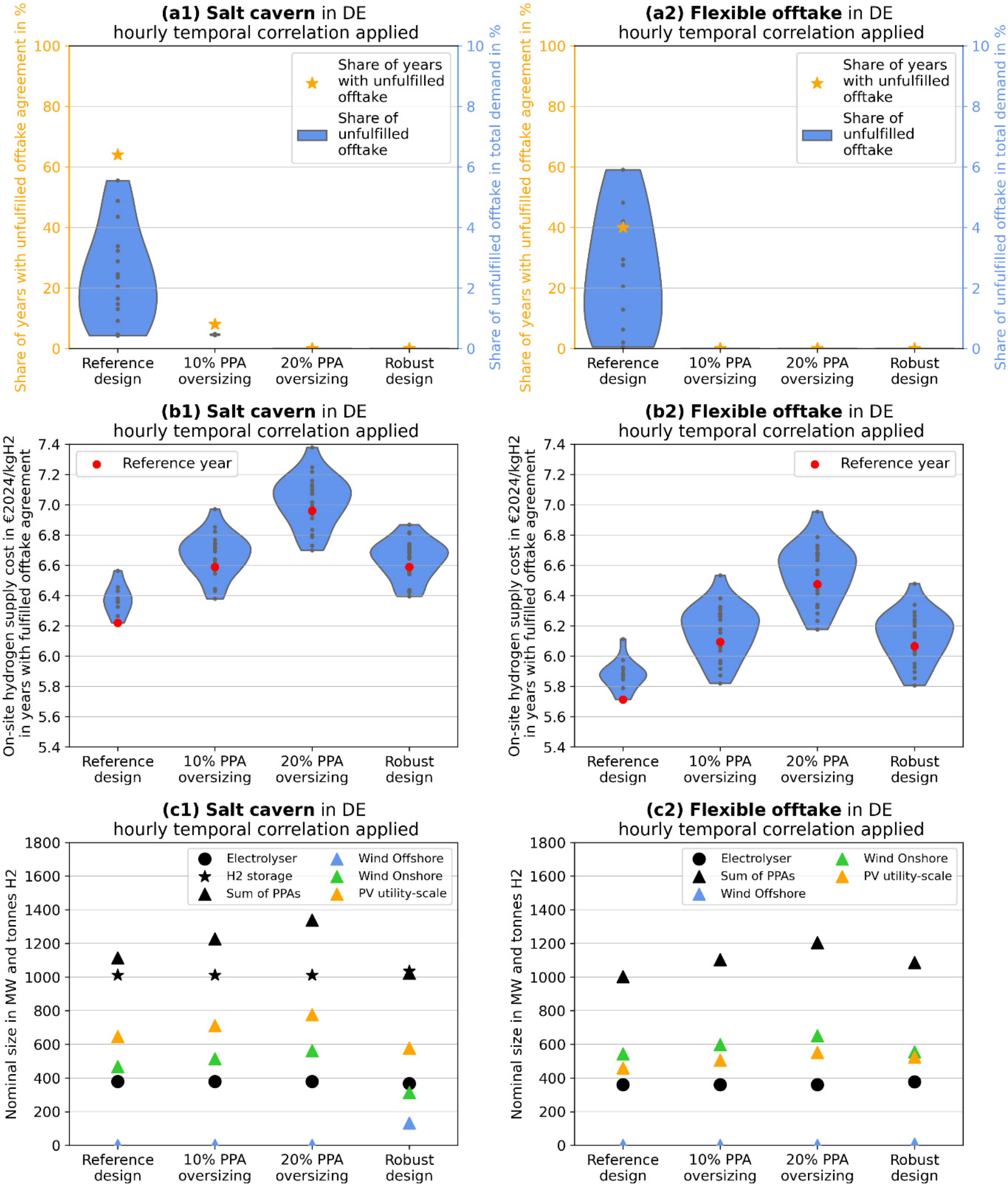

*Figure 9: Comparison of optimisation results between the standard configuration of the considered green hydrogen production system with a salt cavern storage option and a flat offtake characteristic (a1-c1) and a time flexible offtake where no hydrogen storage is needed (a2-c2) for all analysed design paradigms. (a1) and (a2), yellow y-axis: Share of years with unfulfilled offtake agreement out of all analysed years; blue y-axis: Share of total demand needed to meet unfulfilled offtake agreement in years with unfulfilled offtake agreement. (b1) and (b2), On-site hydrogen supply cost in years with fulfilled offtake agreement. (c1) and (c2), Nominal sizes of electrolyser, storage and PPAs. The bidding zone is the German (DE) bidding zone.*

To explore this aspect further, increased offtake flexibility is implemented in two stages. First, the standard configuration of the green hydrogen production system is modified by removing all time restrictions on hydrogen delivery. Thus, only a fixed yearly amount is specified for delivery. Equation (38) shows the respective modification to the mathematical description of the optimisation problem. Second, instead of selling surplus electricity to the electricity market, as investigated in Section 3.1, the PPA surplus is used to produce additional hydrogen by increasing the utilisation of the electrolyser during part-load periods, resulting in a pay-as-produced offtake with a guaranteed annual delivery amount.

Figure 9 shows the same as Figure 5, except (a2), (b2), and (c2) show the results for increased offtake flexibility instead of a flat offtake. Since removing all restrictions on delivery times eliminates the need for hydrogen storage, Figure 9 (c2) does not show storage sizing.

Similar to the regulatory measure explored in the previous Section, increasing offtaker flexibility increases the effectiveness of PPA oversizing in addressing the PPA power volume issue and resulting unfulfilled offtake agreements (see Figure 9 (a2)). As before, 10 % PPA upsizing is sufficient in this regard. In contrast, the maximum share of unfulfilled offtake agreements is slightly increased relative to the flat offtake characteristic in Figure 9 (a1) to 5.9 %. The share of years with unfulfilled offtake agreements, however, is reduced to 40 %. Regarding the *OHSC*, a reduction of 0.49 - 0.52 €/kgH2 is achieved by increased offtake flexibility (see Figure 9 (b1) and (b2)), which, except for the robust design paradigm, exceeds the reductions from relaxed temporal correlation. The *OHSC* spread at 20 % PPA oversizing is the overall maximum value found, at 0.78 €/kgH2. Besides being completely independent of storage capacity across all design paradigms, another distinctive sizing feature is the shift in the PPA portfolio composition towards onshore wind (see Figure 9 (c2)). Cost-optimal robustness is achieved through an even distribution of PPA sizing, with marginal wind offshore integration.

Figure 10 depicts the results of stage two of increased offtake flexibility. (a) shows the additional produced hydrogen from increased electrolyser utilisation through surplus integration. The pale coloured violin plots in (b) show the *OHSC* as in Figure 9 (b2), the dark coloured ones show the corrected *OHSC* after considering the additional hydrogen production. (c) shows the *OHSC* as a function of the surplus selling price, similar to Figure 4, except now only the leftover surplus after additional hydrogen production is sold to the electricity market. Calculation details are in Section 6.3, equations (40), (41), (44) and (45).

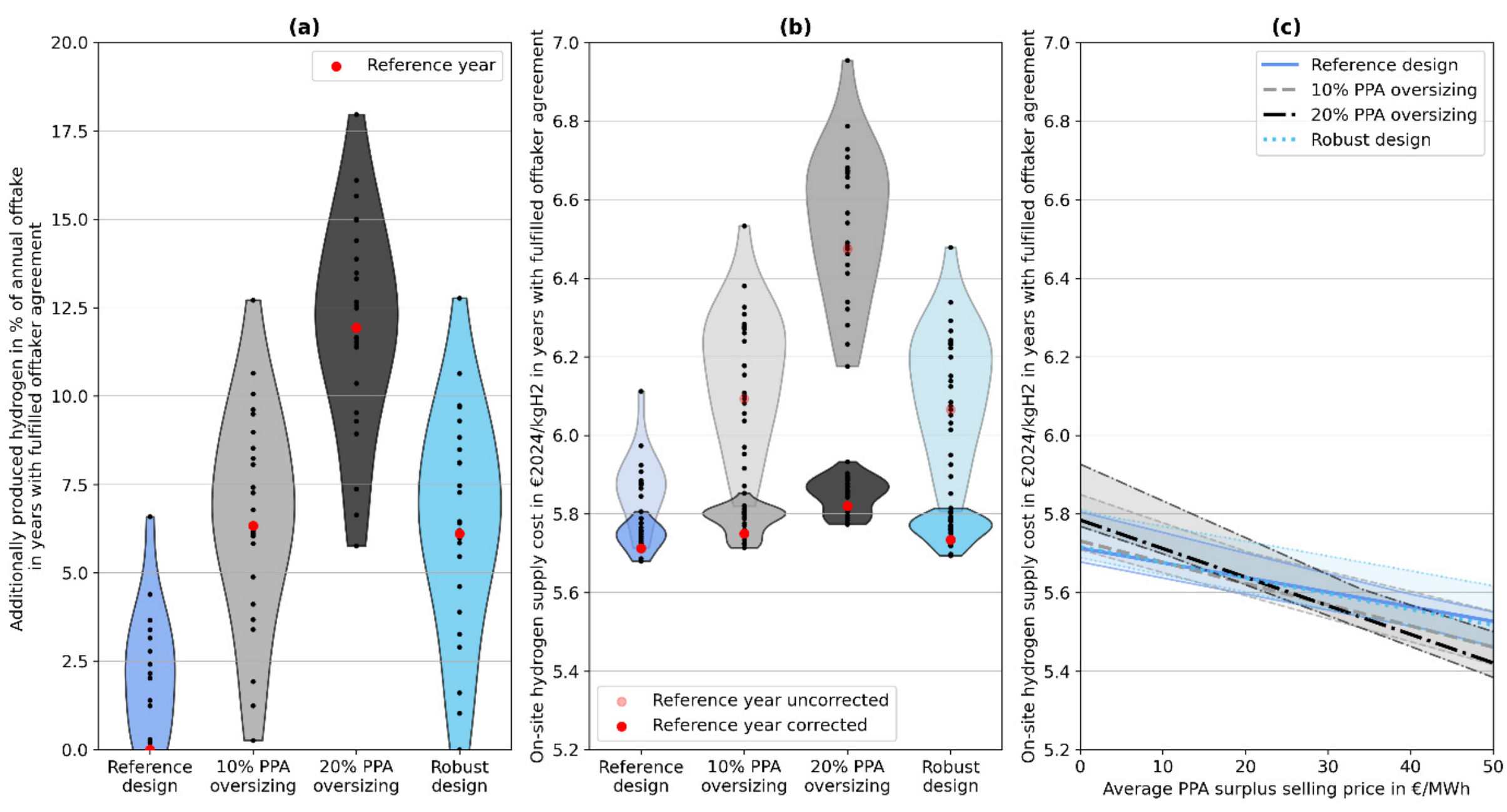

*Figure 10: Optimisation results for the modified green hydrogen production system with a time flexible offtake where no hydrogen storage is needed under all analysed design paradigms with the additional use of surplus PPA power for combined green hydrogen production and PPA surplus selling. (a), Additionally produced hydrogen from PPA surplus use in % of annual offtake in years with fulfilled offtake agreement. (b), Dark coloured violin plots: On-site hydrogen supply cost in years with fulfilled offtake agreement corrected by additionally supplied hydrogen from surplus use. Pale coloured violin plots: On-site hydrogen supply cost in years with fulfilled offtake agreement uncorrected. (c), On-site hydrogen supply cost in years with fulfilled offtake agreement corrected by additional hydrogen production and PPA surplus selling revenues as a function of the PPA surplus selling price. Thick lines within coloured areas in (c) show reference-year values, thin lines on area boundaries show maximum and minimum values. The storage option is a salt cavern storage bundle. The bidding zone is Germany (DE). The temporal correlation is hourly. The hydrogen offtake characteristic is flat and additional hydrogen produced from PPA surplus use on top.*

Using surplus energy for additional hydrogen production can increase the total amount of produced hydrogen by up to 6.6 % under the reference design and by up to 18 % under the 20 % PPA oversizing paradigm, as shown in Figure 10 (a). Assuming the offtaker is willing to take all additional hydrogen, the *OHSC* and its spread can be notably decreased, which can almost offset the *OHSC* increase caused by measures to improve robustness under the PPA oversizing and robust design paradigms (see Figure 10 (b)). Revenues from selling the leftover of surplus energy to the electricity market allow reaching *OHSC* below 5.7 €/kgH2 across all design paradigms at a 50 €/MWh annual average selling price, as shown in Figure 10 (c). Instead of selling the additional hydrogen to the offtaker, it could be stored and used to help fulfil offtake agreements in years with PPA underproduction. Comparing the additional hydrogen production in years with fulfilled offtake agreement under the reference design paradigm in Figure 10 (a) to the shares of unfulfilled offtake in Figure 9 (a2), shows that moving additionally produced hydrogen to years with hydrogen underproduction can help to mitigate the volume risk associated with long-term renewable PPA contracts. However, this requires additional, costly storage upsizing and a certain succession of weather years, avoiding accumulation of hydrogen underproduction.

The same analysis was repeated for the Spanish bidding zone (ES). The results show similar trends with stronger positive effects of increased offtake flexibility on system robustness and lower achievable *OHSC*. The detailed results are in Supplementary Note 4 in the supplementary data.

Overall, increased offtake flexibility can decisively help mitigate the volume risk associated with inter-year fluctuations in long-term PPA contracts resulting from the additionality criterion and can simultaneously lead to notable *OHSC* reductions. Maximal offtake flexibility in a pay-as-produced contract can compensate for additional costs from operator design measures needed to improve the robustness of green hydrogen production systems and can narrow the *OHSC* spread caused by inter-year PPA fluctuations.

# 4 Conclusions and policy implications

According to the European Union regulatory framework, green hydrogen production in the European Union is subject to the rule of additionality, necessitating the use of renewable energy from newly built, additional, energy sources if power is sourced via renewable PPAs. Since newly built renewable PPAs are only bookable as long-term contracts, and temporal correlation rules prohibit short-term compensation via the electricity market, their weather-dependent inter-year production fluctuations pose a volume risk to an electrolyser operator. On the one hand, he is dependent on an unreliable, inter-year fluctuating power source. On the other hand, he has to fulfil a green hydrogen offtake agreement with time- and volume-constraints from major hydrogen users such as refineries, chemical plants, and steel factories.

By applying different design paradigms to a green hydrogen production system operating with renewable PPA power, this work shows that inter-year fluctuations of long-term renewable PPAs can notably affect on-site hydrogen supply cost and jeopardise the fulfilment of offtake agreements. Depending on different modifications of the considered production system, such as hydrogen storage options or the bidding zone in the European Union, in up to 85 % of the 25 analysed weather years, up to 5.9 % of hydrogen demand cannot be delivered due to PPA volume fluctuations. Simple PPA oversizing turned out to be valid method for addressing this issue for some evaluated system modifications. However, it entails additional power sourcing costs of up to 0.96 €/kgH2 for 20 % PPA overbooking and increases inter-year variations in supply costs to up to 0.78 €/kgH2 between high- and low-weather years. Revenues from selling surplus electricity can offset the additional cost and inter-year cost variations from PPA upsizing. Whereas the effectiveness depends on electricity market prices. In contrast to simple PPA upsizing, PPA diversification, combined with storage capacity upsizing is the most cost-effective electrolyser operator measure to increase the robustness of green hydrogen production systems.

As an alternative, offtake flexibility could play a crucial role in solving the volume issue posed by the additionality rule. Time-flexible offtake profiles notably reduce the risk of unfulfilled offtake agreements due to inter-year volume fluctuations in renewable PPA feed-in. If green hydrogen offtakers are flexible enough to take hydrogen on a pay-as-produced basis, the magnitude and spread of production costs are notably reduced, and additional costs from electrolyser operator measures to increase system robustness, such as PPA or storage upsizing, can be offset. Despite relatively flat production profiles in industrial green hydrogen applications, increased offtake flexibility is imaginable, especially in the coming years when green hydrogen blending is possible, and grey hydrogen supply flexibility is available. As soon as a short-term green hydrogen market is established, additional flexibilities could arise. However, increasing green hydrogen application quota in the future could have opposing effects.

Finally, the transitional rule for temporal correlation, which allows limited electricity market-based compensation for first-mover projects until 2030, notably decreases the probability of unfulfilled offtake agreements and reduces green hydrogen supply costs by up to 0.6 €/kgH2. However, the impact magnitude of a relaxed temporal correlation rule on system robustness and production costs depends on grid electricity prices and the quality of renewable PPAs available in the different European bidding zones. The same accounts for the increased emission intensity, resulting from grid electricity integration. Whereas the emission intensity exceeds the minimum emission-saving threshold for green hydrogen only at electricity prices below 50 €/MWh, in single years, and in bidding zones with comparably high emission intensity levels of grid electricity. In summary, the transitional rule for temporal correlation provides a good compromise between robustness increase, production cost decrease and emission avoidance. Thus, extending the transitional rules for temporal correlation would reduce the volume risk posed by the additionality criterion. If addressed by the European Commission when reviewing the power sourcing criteria in 2026, this could provide a boost to renewable PPA-based green hydrogen production, help close the widening implementation gap and thereby reducing Europe's dependence on fossil energy imports. In addition, ensuring a level playing field with other possible green and low-carbon hydrogen production routes. However, for an extension to be effective in terms of production system robustness, its duration would have to match the long-term duration of renewable PPA contracts, which strengthens the call for temporal correlation rule adjustments proposed by ministries from several European Union member states in early 2026 [5].

## 5 Limitations and outlook

We explicitly emphasise that the conclusions drawn in this study are based on the assumptions made. Critical aspects here are the limited geographical diversification of PPA options, which allows us to make technology-specific findings and the simplification of pricing and bidding mechanisms through the assumption of fixed annual average prices when electricity mix power sourcing is allowed. The same applies to the assumption of annual average values of electricity emission intensity for emission calculations, as introduced by the currently valid European Commission methodology. Since the occurrence probability of over- and under-producing weather years may affect the cost-optimal design and operation of green hydrogen production systems, their integration through stochastic optimisation could yield additional insights into the explored subject in subsequent studies. The same applies to further diversification of offtake flexibility, where various application-specific variants may occur.

Due to the complexity of the subject and the limitations mentioned, this study alone cannot provide an all-encompassing picture of the effects of the European Union's power purchase criteria for green hydrogen production. Nevertheless, its findings should specifically contribute to a better understanding of how the additionality criterion and associated long-term contracts for renewable PPAs affect green hydrogen production costs and the fulfilment of offtake agreements. Furthermore, insights into how offtake flexibility and regulatory adjustments can help mitigate the volume risk posed by the additionality criterion should provide valuable information to policymakers, economic players and especially regulators in the upcoming review of the power sourcing criteria for green hydrogen production.

# 6 Appendix

## 6.1 Appendix A. Mathematical description of optimisation problems

### 6.1.1 Reference design optimisation

To optimise design and operation of the chosen system modification in the reference design optimisation (see Figure 2 (a) step 1.), the total annual cost of the system $C_{\mathrm{tot}}$ is minimised (compare with equation (1)). It consists of the sum of annualized capital cost $C_{\mathrm{TAC,CAPEX},i}$ and total annual operation cost $C_{\mathrm{TAC,OPEX},i}$ for each component $i$ in the hydrogen production plant and the annual power purchase cost $C_{\mathrm{TAC,PP,j}}$ for each power purchase option $j$ in Figure 1.

$$C_{\mathrm{tot}} = \sum_{i=1}^{N} (C_{\mathrm{TAC,CAPEX},i} + C_{\mathrm{TAC,OPEX},i}) + \sum_{j=1}^{M} C_{\mathrm{TAC,PP},j} \tag{3}$$

The annualized capital cost of each component in the hydrogen production plant is calculated by multiplying the nominal power $P_{\mathrm{nom},i}$ by the specific capital cost $c_{\mathrm{CAPEX},i}$ and the annuity payment factor $A_i$. Equation (5) shows the equivalent for hydrogen storage components.

$$C_{\mathrm{TAC,CAPEX},i} = P_{\mathrm{nom},i} \cdot c_{\mathrm{CAPEX},i} \cdot A_i \tag{4}$$

$$C_{\mathrm{TAC,CAPEX,Sto}} = m_{\mathrm{nom,Sto}} \cdot c_{\mathrm{CAPEX,Sto}} \cdot A_{\mathrm{Sto}} \tag{5}$$

The annuity payment factor is calculated as follows, where $r_{\mathrm{in},i}$ is the interest rate and $t_{\mathrm{dep},i}$ the depreciation time of the respective component. Hereby, the interest rate displays the nominal weighted average cost of capital.

$$A_i = \frac{r_{\mathrm{in},i} \cdot \left(1 + r_{\mathrm{in},i}\right)^{t_{\mathrm{dep},i}}}{\left(1 + r_{\mathrm{in},i}\right)^{t_{\mathrm{dep},i}} - 1} \tag{6}$$

The annual operational cost of each component in the hydrogen production plant is calculated by multiplying the nominal power $P_{\mathrm{nom},i}$ by the specific capital cost $c_{\mathrm{CAPEX},i}$ and a fixed operational cost factor $f_{\mathrm{OPEXfix},i}$. To this fixed operational cost part a variable part is added by multiplying the annual sum of purchased power $P_{i,t}$ by the time step length and a variable operational cost factor $f_{\mathrm{OPEXvar},i}$, as shown in equation (7). Equation (8) shows the equivalent for hydrogen storage components.

$$C_{\mathrm{TAC,OPEX},i} = P_{\mathrm{nom},i} \cdot c_{\mathrm{CAPEX},i} \cdot f_{\mathrm{OPEXfix},i} + \sum_{t=1}^{T} P_{i,t} \cdot \Delta t \cdot f_{\mathrm{OPEXvar},i} \tag{7}$$

$$C_{\mathrm{TAC,OPEX,Sto}} = m_{\mathrm{nom,Sto}} \cdot c_{\mathrm{CAPEX,Sto}} \cdot f_{\mathrm{OPEXfix,Sto}} + \sum_{t=1}^{T} \dot{m}_{\mathrm{StoIn},t} \cdot \Delta t \cdot f_{\mathrm{OPEXvar,Sto}} \tag{8}$$

The total cost for power purchase for power purchase option $j$ is calculated by multiplying the sum of purchased power $P_{j,t}$ of power purchase option $j$ for all time steps $t$ in the optimisation time frame {1,2,3,…, $T$} by the length of a time step $\Delta t$ and the specific power purchase price $p_j$.

$$C_{\mathrm{TAC,PP},j} = \sum_{t=1}^{T} P_{j,t} \cdot \Delta t \cdot p_j \tag{9}$$

In case of PPAs, $P_{j,t}$ is replaced by the multiplication of the nominal power of the purchased PPA option $P_{\mathrm{nom},j}$ and the capacity factor $f_{\mathrm{cap},j,t}$ of the PPA option at time step $t$.

$$C_{\mathrm{TAC,PP},j} = \sum_{t=1}^{T} P_{\mathrm{nom},j} \cdot f_{\mathrm{cap},j,t} \cdot \Delta t \cdot p_j \tag{10}$$

The following equations are the equality constraints that define the technical operation of the optimized hydrogen production system in Figure 1**Fehler! Verweisquelle konnte nicht gefunden werden.**. All optimisation variables and parameters and their descriptions are listed in

Table *2* and

Table *3*.

$$P_{\mathrm{RD},t} + P_{\mathrm{WindOff},t} + P_{\mathrm{WindOn},t} + P_{\mathrm{PV},t} + P_{\mathrm{Grid},t} = P_{\mathrm{Ely},t} + P_{\mathrm{Comp},t} \qquad \forall t \epsilon \{1,2,3,\dots,T\} \tag{11}$$

$$P_{\mathrm{Ely},t} = P_{\mathrm{Peri},t} + P_{\mathrm{Stack},t} \qquad \forall t \epsilon \{1,2,3,\dots,T\} \tag{12}$$

$$P_{\mathrm{ElySys},t} = P_{\mathrm{Ely},t} + P_{\mathrm{Comp},t} \qquad \forall t \epsilon \{1,2,3,\dots,T\} \tag{13}$$

$$P_{\mathrm{Peri},t} = \dot{m}_{\mathrm{Ely},t} \cdot \varepsilon_{\mathrm{Peri},t} \qquad \forall t \epsilon \{1,2,3,\dots,T\} \tag{14}$$

$$P_{\mathrm{Comp},t} = \dot{m}_{\mathrm{Ely},t} \cdot \varepsilon_{\mathrm{Comp},t} \qquad \forall t \epsilon \{1,2,3,\dots,T\} \tag{15}$$

$$\dot{m}_{H_2O,t} = \dot{m}_{\mathrm{Ely},t} \cdot \varepsilon_{\mathrm{H_2O},t} \qquad \forall t \epsilon \{1,2,3,\dots,T\} \tag{16}$$

$$\dot{m}_{\mathrm{Ely},t} \cdot \left(1 - f_{\mathrm{loss,Comp}}\right) = \dot{m}_{\mathrm{StoIn},t} - \dot{m}_{\mathrm{StoOut},t} + \dot{m}_{\mathrm{demand},t} \qquad \forall t \epsilon \{1,2,3,\dots,T\} \tag{17}$$

$$m_{\mathrm{Sto},t=1} = m_{\mathrm{Sto},t=\mathrm{T}} + \left(\dot{m}_{\mathrm{StoIn},t=1} - \dot{m}_{\mathrm{StoOut},t=1}\right) \cdot \Delta t \qquad \forall t \epsilon \{1,2,3,\dots,T\} \tag{18}$$

$$m_{\mathrm{Sto},t} = m_{\mathrm{Sto},t-1} + \left(\dot{m}_{\mathrm{StoIn},t} - \dot{m}_{\mathrm{StoOut},t}\right) \cdot \Delta t \qquad \forall t \epsilon \{1,2,3,\dots,T\} \quad (19)$$

The following equations are the inequality constraints that connect the operation and design of all relevant components and power purchase options.

$$P_{\mathrm{WindOff},t} \leq P_{\mathrm{nom,WindOff}} \cdot f_{\mathrm{cap,WindOff},t} \qquad \forall t \epsilon \{1,2,3,\dots,T\} \quad (20)$$

$$P_{\mathrm{WindOn},t} \leq P_{\mathrm{nom,WindOn}} \cdot f_{\mathrm{cap,WindOn},t} \qquad \forall t \epsilon \{1,2,3,\dots,T\} \quad (21)$$

$$P_{\mathrm{PV},t} \leq P_{\mathrm{nom,PV}} \cdot f_{\mathrm{cap,PV},t} \qquad \forall t \epsilon \{1,2,3,\dots,T\} \quad (22)$$

$$P_{\mathrm{RD},t} \leq P_{\mathrm{RDava},t} \qquad \forall t \epsilon \{1,2,3,\dots,T\} \quad (23)$$

$$P_{\mathrm{Ely},t} \leq P_{\mathrm{nom,Ely}} \qquad \forall t \epsilon \{1,2,3,\dots,T\} \quad (24)$$

$$P_{\mathrm{Stack},t} \leq P_{\mathrm{nom,Stack}} \qquad \forall t \epsilon \{1,2,3,\dots,T\} \quad (25)$$

$$P_{\mathrm{Peri},t} \leq P_{\mathrm{nom,Peri}} \qquad \forall t \epsilon \{1,2,3,\dots,T\} \quad (26)$$

$$P_{\mathrm{Comp},t} \leq P_{\mathrm{nom,Comp}} \qquad \forall t \epsilon \{1,2,3,\dots,T\} \quad (27)$$

$$m_{\mathrm{Sto},t} \leq m_{\mathrm{nom,Sto}} \qquad \forall t \epsilon \{1,2,3,\dots,T\} \quad (28)$$

To map the decrease in specific energy consumption of the electrolyser stack at operating points below the nominal load while keeping the mathematical formulation of the optimisation problem linear, the linearization method as in [9] was used. The respective constraints to construct a convex search space to keep the optimisation problem linear are the following.

$$\dot{m}_{\mathrm{Ely},t} \leq a_{\mathrm{lin},k} \cdot P_{\mathrm{Stack},t} + b_{\mathrm{lin},k} \qquad \forall t \epsilon \{1,2,3,\dots,T\} \text{ and } \forall k \epsilon \{0,1,2,\dots,K-1\} \quad (29)$$

with

$$a_{\mathrm{lin},k} = \frac{k+1}{\varepsilon_{\mathrm{Stack},k+1}} - \frac{k}{\varepsilon_{\mathrm{Stack},k}} \quad (30)$$

$$b_{\mathrm{lin},k} = \frac{\frac{k}{K-1} \cdot P_{\mathrm{nom,Stack}}}{\varepsilon_{\mathrm{Stack},k}} - a_{\mathrm{lin},k} \cdot \frac{k}{K-1} \cdot P_{\mathrm{nom,Stack}} \quad (31)$$

Hereby, $b_{\mathrm{lin},k}$ is the y-axis intersect of the respective linear constraint $k$. The number of total linearization steps $K$ defines the number of additional inequality constraints needed per time step $t$. The number of linearization steps chosen in this study is 20. The inequality constraint in Equation (32) defines the lower bound of the constructed search space and is used to reduce the optimisation time by minimising the search space.

$$P_{\mathrm{Stack},t} \leq \dot{m}_{\mathrm{Ely},t} \cdot \varepsilon_{\mathrm{nom,Stack}} \qquad \forall t \epsilon \{1,2,3,\dots,T\} \quad (32)$$

*Table 2: Optimisation variables*

| **Variable** | **Description** |
| --- | --- |
| $m_{\text{nom,Sto}}$ | Nominal capacity of hydrogen storage |
| $m_{\text{Sto},t}$ | Stored hydrogen mass at time step $t$ |
| $\dot{m}_{\text{Ely},t}$ | Hydrogen output mass flow of electrolyser plant at time step $t$ |
| $\dot{m}_{\text{StoIn},t}$ | Hydrogen mass flow into hydrogen storage at time step $t$ |
| $\dot{m}_{\text{StoOut},t}$ | Hydrogen mass flow out of hydrogen storage at time step $t$ |
| $P_{\text{nom,Comp}}$ | Nominal power of compressor |
| $P_{\text{Comp},t}$ | Compressor power consumption at time step $t$ |
| $P_{\text{nom,Ely}}$ | Nominal power of electrolyser |
| $P_{\text{Ely},t}$ | Electrolyser power consumption at time step $t$ |
| $P_{\text{ElySys},t}$ | Electrolyser system (including compressor) power consumption at time step $t$ |
| $P_{\text{Grid},t}$ | Grid power supply at time step $t$. If hourly temporal correlation applies, $P_{\text{Grid},t} = 0 \;\; \forall t \epsilon \{1,2,3, \dots, \text{T}\}$. |
| $P_{\text{nom,Peri}}$ | Nominal power of electrolyser peripherals |
| $P_{\text{Peri},t}$ | Electrolyser peripherals power consumption at time step $t$ |
| $P_{\text{nom,PV}}$ | Nominal power of photovoltaic |
| $P_{\text{PV},t}$ | Photovoltaic power at time step $t$ |
| $P_{\text{nom,Stack}}$ | Nominal power of electrolyser stack |
| $P_{\text{Stack},t}$ | Electrolyser stack power consumption at time step $t$ |
| $P_{\text{nom,WindOff}}$ | Nominal power of wind offshore turbine |
| $P_{\text{WindOff},t}$ | Wind offshore turbine power at time step $t$ |
| $P_{\text{nom,WindOn}}$ | Nominal power of wind onshore turbine |
| $P_{\text{WindOn},t}$ | Wind onshore turbine power at time step $t$ |

*Table 3: Optimisation parameters*

| **Parameter** | **Description** |
|---|---|
| $a_{\text{lin},k}$ | Gradient of linearized characteristics curve of electrolyser stack through linearization steps $k$ and $k + 1$ |
| $b_{\text{lin},k}$ | The y-axis intersect of linearized characteristics curve of electrolyser stack through linearization steps $k$ and $k + 1$ |
| $f_{\text{cap,PV},t}$ | Photovoltaic capacity factor at time step $t$ |
| $f_{\text{cap,WindOff},t}$ | Wind offshore capacity factor at time step $t$ |
| $f_{\text{cap,WindOn},t}$ | Wind onshore capacity factor at time step $t$ |
| $f_{\text{loss,Comp}}$ | Hydrogen loss factor of compressor |
| $\dot{m}_{\text{demand},t}$ | Hydrogen demand mass flow (predefined) at time step $t$. As mentioned in Section 2.1, the annual sum of green hydrogen demand is $36{,}500\ \text{tonnes H2}$. With a flat distribution, at each time step $t$ an hourly demand of $4.16\overline{7}\,\frac{\text{kgH2}}{\text{h}}$ results. |
| $\Delta t$ | Length of time step in hours |
| $\varepsilon_{\text{Comp}}$ | Specific energy consumption of compressor |
| $\varepsilon_{\text{H}_2\text{O}}$ | Specific water consumption of electrolyser plant |
| $\varepsilon_{\text{Peri}}$ | Specific energy consumption of electrolyser peripherals |
| $\varepsilon_{\text{nom,Stack}}$ | Specific energy consumption of electrolyser stack at nominal power |
| $\varepsilon_{\text{Stack},k}$ | Specific energy consumption of electrolyser stack at $\frac{k}{K-1}\cdot 100\ (\%)$ of nominal power |

In order to evaluate the performance of the reference design in the different weather years, first the described optimisation problem is solved, using the capacity factor time series from the reference year as an input ((see Figure 2 (a) step 1.). Then, two changes compared to the reference design optimisation are made ((see Figure 2 (a) step 2.):

1. All nominal design variables from Table 2 are set to design values resulting from the reference design optimisation in the reference weather year.
2. The capacity factor time series of the renewable PPAs in (see
3. Table *3*) are replaced by the time series of all other analysed weather years.

The resulting optimisation, which only minimises the operational cost of the green hydrogen production system, is then performed for all 25 considered weather years individually.

### 6.1.2 PPA oversizing optimisation

To perform the PPA oversizing, the nominal values of all PPA design variables resulting from the reference design optimisation are multiplied by an oversizing factor $f_{\text{oversize}}$ (see Figure 2 (b) 1.). Equation (33) shows the upsizing, exemplary for the utility-scale photovoltaic PPA.

$$P_{\text{nom,PV}}^{\text{PPAover}} = P_{\text{nom,PV}}^{\text{Ref}} \cdot f_{\text{over}} \tag{33}$$

The further procedure for performance evaluation of the PPA oversizing is the same as in Section 6.1.1 for the reference design performance evaluation, but now the nominal PPA design values from the reference design optimisation are replaced by the oversized values. Then, they are used as a precondition besides all other nominal values from the reference design optimisation (see Figure 2 (b) 2.) to perform the 25 operational optimisations individually. The considered oversizing factors for 10 % and 20 % oversizing are 1.1 and 1.2.

### 6.1.3 Robust design optimisation

The goal of robust design optimisation is finding a system design for the considered hydrogen production system that ensures offtake fulfilment in all considered years at minimised cost. The used method for robust design optimisation is oriented on the concept of “optimality robustness” [28]. Applying this concept to the optimisation problem described in Section 6.1.1 and considering the renewable PPA feed-in time series of the analysed weather years, the robust optimisation problem consists of $S$ sub-problems, one for each weather year (see Figure 11).

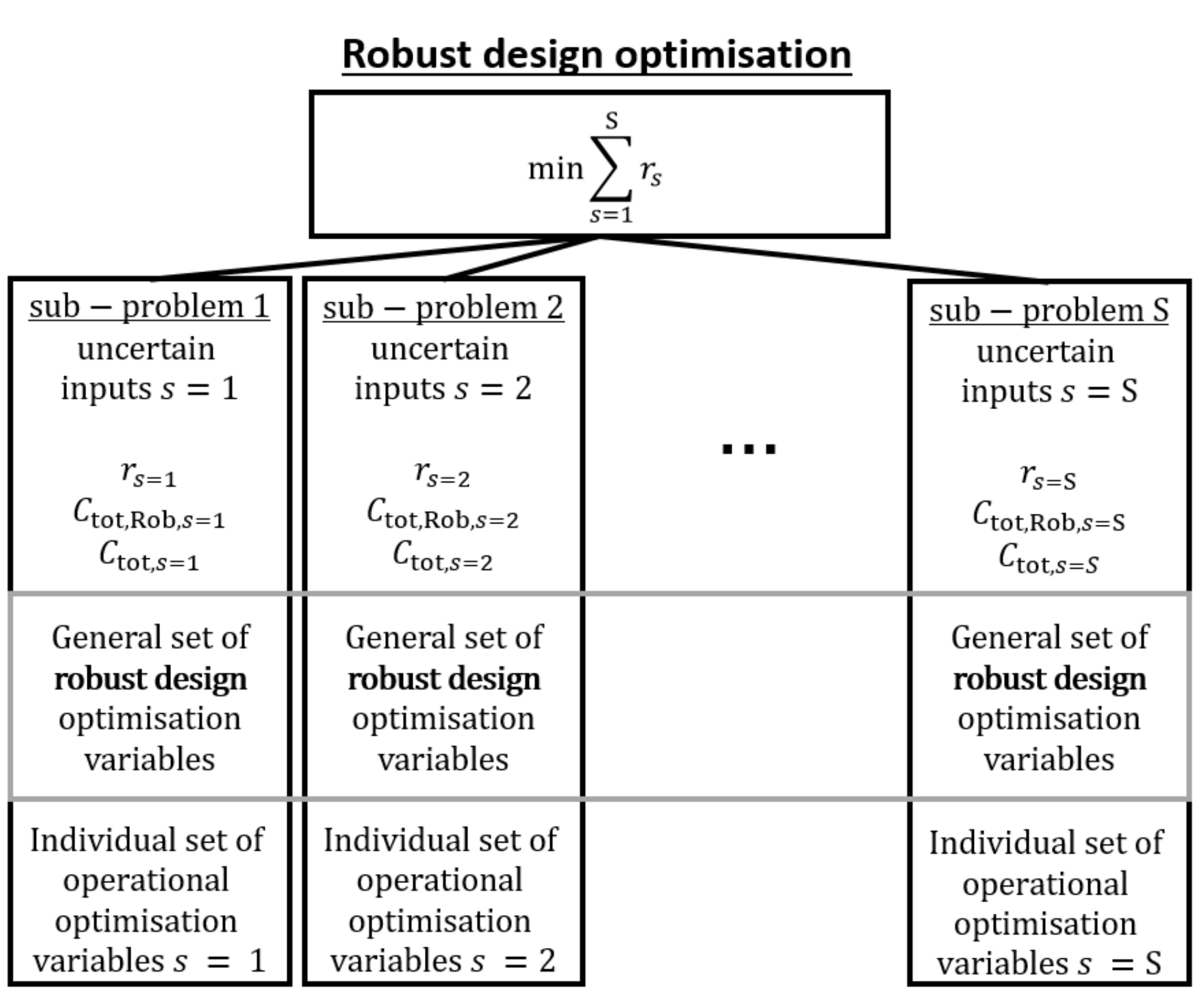

*Figure 11: Graphical display of robust design optimisation.*

Each sub-problem $s$ has an individual set of uncertain inputs, in this case the capacity-factor time series of each weather year for all considered renewable PPA options. Furthermore, the operational variables in each sub-problem are sub-problem specific. In contrast, the design variables of the considered system are the same for all sub-problems and thereby serve as connecting element between the sub-problems. For each sub-problem the total cost $C_{\text{tot,Rob},s}$ (see equation (34) and compare with equation (3)) result.

$$C_{\text{tot,Rob},s} = \sum_{i=1}^{N} (C_{\text{TAC,CAPEX},i} + C_{\text{TAC,OPEX},i,s}) + \sum_{j=1}^{M} C_{\text{TAC},PP,j,s} \qquad \forall s \epsilon \{0,1,2,\dots,S\} \quad (34)$$

The annual operation cost $C_{\text{TAC,OPEX},i,s}$ and the annual power purchase cost $C_{\text{TAC,PP},j,s}$ differ between the sub-problems. The sum of annualized capital cost $C_{\text{TAC,CAPEX},i}$ are the same in all years due to uniform sizing in all sub-problems.

To perform robust design optimisation for each sub-problem $s$ a regret value $r_s$ is calculated by subtracting the total cost occurring in the sub-problem $s$ of the robust optimisation $C_{\text{tot,Rob},s}$ from the total cost $C_{\text{tot},s}$ (see equation (35)).

$$r_s = C_{\text{tot},s} - C_{\text{tot,Rob},s} \qquad \forall s \epsilon \{0,1,2,\dots,S\} \quad (35)$$

$C_{\text{tot},s}$ serves as a reference for the minimal achievable cost in year $s$ if singularly evaluated and is provided by performing the reference design optimisation described in Section 6.1.1 for the weather year $s$ (see equation (3)).

In contrast to the reference design optimisation in Section 6.1.1, here the sum of regrets is minimised (see equation (36)) to achieve cost-minimised robustness.

$$\min \sum_{s=1}^{S} r_s \tag{36}$$

## 6.2 Appendix B. System modifications

Equation (37) shows the additional constraints for the relaxed temporal correlation condition from hourly to monthly correlation, analysed in Section 3.4. Hereby the sum of electrolyser system power in month $l$ is not allowed to exceed the sum of contracted PPA power in the respective month.

$$\sum_{t=l\cdot 730+1}^{(l+1)\cdot 730} P_{\text{ElySys},t} \leq \sum_{t=l\cdot 730+1}^{(l+1)\cdot 730} (P_{\text{nom,WindOff}} \cdot f_{\text{cap,WindOff},t} + P_{\text{nom,WindOn}} \cdot f_{\text{cap,WindOn},t} + P_{\text{nom,PV}} \cdot f_{\text{cap,PV},t}) \quad \forall l \epsilon \{0,1,2,\ldots,11\} \tag{37}$$

Equation (38) shows the modification of equation (17), allowing the analyses of increased offtake flexibility in Section 3.5. The annual sum of produced hydrogen is equal to the annual sum of hydrogen demand.

$$\sum_{t=1}^{T} (\dot{m}_{\text{Ely},t} \cdot (1 - f_{\text{loss,Comp}})) = \sum_{t=1}^{T} \dot{m}_{\text{demand},t} \tag{38}$$

## 6.3 Appendix C. Evaluation indices

The share of annual surplus in total PPA production $S_{\text{surplus,PPA}}$ (see Figure 4 (a)) is calculated by dividing the difference between total annual PPA production potential and actual total annual PPA production by the total annual PPA production potential (see equation (39)).

$$S_{\text{surplus,PPA}} = \frac{\sum_{t=1}^{T} \left( (P_{\text{nom,WindOff}} \cdot f_{\text{cap,WindOff},t} + P_{\text{nom,WindOn}} \cdot f_{\text{cap,WindOn},t} + P_{\text{nom,PV}} \cdot f_{\text{cap,PV},t}) - (P_{\text{WindOff},t} + P_{\text{WindOn},t} + P_{\text{PV},t}) \right)}{\sum_{t=1}^{T} (P_{\text{nom,WindOff}} \cdot f_{\text{cap,WindOff},t} + P_{\text{nom,WindOn}} \cdot f_{\text{cap,WindOn},t} + P_{\text{nom,PV}} \cdot f_{\text{cap,PV},t})} \tag{39}$$

The corrected on-site hydrogen supply cost from surplus selling $OHSC_{\text{corr,surplus}}$ (see Figure 4 (c)) is calculated by subtracting the revenues from surplus selling divided by the sum of annual hydrogen production from the $OHSC$ calculated after optimisation (see equation (2)) as shown in equation (40). The revenues from surplus selling are calculated by multiplying the annual sum of surplus energy by the average surplus selling price $p_{\text{av,sell}}$ (see equation (41)).

$$OHSC_{\text{corr,surplus}} = OHSC - \frac{R_{\text{surplus}}}{\sum_{t=1}^{\text{T}} \dot{\text{m}}_{\text{demand},t}} \tag{40}$$

$$R_{\text{surplus}} = \sum_{t=1}^{T} \Big( \big( P_{\text{nom,WindOff}} \cdot f_{\text{cap,WindOff},t} + P_{\text{nom,WindOn}} \cdot f_{\text{cap,WindOn},t} + P_{\text{nom,PV}} \cdot f_{\text{cap,PV},t} \big) - \big( P_{\text{WindOff},t} + P_{\text{WindOn},t} + P_{\text{PV},t} \big) \Big) \cdot p_{\text{av,sell}} \tag{41}$$

The emission intensity of produced hydrogen $i_{\text{em,H2}}$ (see Figure 8 (c)) is calculated by dividing the product of annually used grid electricity and the average grid emission intensity $i_{\text{em,av,Grid}}$ by the annual amount of produced hydrogen (see equation (42)). The use of the average emission intensity values of grid electricity is oriented on the calculation methodology for greenhouse gas emission savings from green hydrogen, introduced by the European Commission [37].

$$i_{\text{em,H2}} = \frac{\sum_{t=1}^{T} P_{\text{Grid},t} \cdot i_{\text{em,av,Grid}}}{\sum_{t=1}^{T} \dot{m}_{\text{demand},t}} \tag{42}$$

The utilisation of the electrolyser in annual full load hours $U_{\text{Ely}}$, depicted in Figure 8 (f), is calculated by dividing the annual sum of electrolyser energy consumption by the nominal size of the electrolyser (see equation (43)).

$$U_{\text{Ely}} = \frac{\sum_{t=1}^{T} P_{\text{Ely},t} \cdot \Delta t}{P_{\text{nom,Ely}}} \tag{43}$$

To be able to calculate the additionally produced hydrogen depicted in Figure 10 (a), the power consumption of the electrolyser $P_{\text{Ely},t}$ resulting from dispatch and design optimisation is increased at times where it was below its nominal value $P_{\text{nom,Ely}}$ and in case surplus power from the PPAs is available. By applying equations (12)-(15) and (29) and the hydrogen loss factor of the compressor, the modified power consumption of the electrolyser $P_{\text{Ely},t,\text{mod}}$ was then used to calculate a modified hydrogen production time series $\dot{m}_{\text{Ely},t,\text{mod}}$. By dividing the difference of its annual sum and the annual sum of predefined hydrogen demand by the annual sum of predefined hydrogen demand, the additionally produced hydrogen $S_{\text{add,H2}}$ as a share of the total annual demand, depicted in Figure 10 (a), is calculated as in equation (44).

$$S_{\text{add,H2}} = \frac{\sum_{t=1}^{T} \dot{m}_{\text{Ely},t,\text{mod}} - \sum_{t=1}^{T} \dot{m}_{\text{demand},t}}{\sum_{t=1}^{T} \dot{m}_{\text{demand},t}} \tag{44}$$

The corrected on-site hydrogen supply cost from additional hydrogen production $OHSC_{\text{corr,add}}$ (see Figure 10 (b)), is calculated by multiplying the $OHSC$ calculated after optimisation (see equation (2)) and the ratio between the predefined annual sum of hydrogen demand and the sum of the modified hydrogen production time series and by adding the cost for additional water purchase for electrolyser operation (see equation (45)).

$$OHSC_{\text{corr,add}} = OHSC \cdot \frac{\sum_{t=1}^{\text{T}} \dot{m}_{\text{demand},t}}{\sum_{t=1}^{\text{T}} \dot{m}_{\text{Ely},t,\text{mod}}} + \frac{\left(\sum_{t=1}^{\text{T}} \dot{m}_{\text{Ely},t,\text{mod}} - \sum_{t=1}^{\text{T}} \dot{m}_{\text{demand},t}\right) \cdot \varepsilon_{\text{H}_2\text{O}} \cdot p_{\text{H}_2\text{O}}}{\sum_{t=1}^{\text{T}} \dot{m}_{\text{Ely},t,\text{mod}}} \tag{45}$$

Further correction of the on-site hydrogen cost $OHSC_{\text{corr,add}}$ by surplus selling, as shown in Figure 10 (c), is performed equally as shown in equation (40). In contrary to the surplus selling correction performed in Figure 4, in this case the surplus energy from PPA production is reduced by the surplus used for additional hydrogen production.

## 6.4 Appendix D. Supplementary data

Supplementary material provided with this paper including techno-economic assumptions, details on PPA modeling and price calculation and additional result figures referenced in Sections 3.4 and 3.5.

# 7 Software

The optimisation problem was implemented in python [39]. Gurobi [40] was used as mathematical solver for all optimisations carried out in this work.

# 8 Data and code availability

All data used, generated or analysed in the course of this study have been deposited at Zenodo [41] and is publicly available at https://zenodo.org/records/20617315.

The mathematical description of the implemented optimisation problems, including the objective function, all constraints and all parameter assumptions made is documented in detail and comprehensively in the central article, Methods and Supplementary data. In addition, all original code has been deposited at Zenodo [41] and is publicly available at https://zenodo.org/records/20617315.

Any additional information required to reanalyse the data reported in this paper is available from the lead contact upon request.

# 9 Acknowledgments

This work was funded by the German BMWE within the SyNerGy-H2 project (grant number 03EI6143A). The results presented were achieved by computations carried out on the cluster system at the Leibniz Universität Hannover, Germany. We want to thank Nicolas Stünkel, Tim Semmler and Joel Horn for their support in computational implementation work done in the course of this study.

# 10 Author contributions

Brandt, Jonathan (Conceptualization, Software, Analysis, Visualization, Methodology, Writing - original draft, Writing – review & editing)

Bensmann, Astrid; Hanke-Rauschenbach, Richard (Conceptualization, Project administration, Supervision, Writing – review & editing, Funding acquisition)

# 11 Corresponding authors

Correspondence to Jonathan Brandt or Astrid Bensmann

# 12 Declaration of interests

The authors declare no competing interests.

# C References

Mitigating business risks from renewable PPA power sourcing uncertainties for European green hydrogen production: Robust system design, regulatory adjustments and offtake flexibility

# Supplementary material

# Supplementary Note 1: Technical and economical parameter assumptions

*Supplementary Table 1: Technical and economical parameter assumptions.*

| Component | Parameter | Value | Unit | Reference |
|---|---|---|---|---|
| Wind offshore PPA (Germany) | Pay-as-produced price* | 88.5 | $€_{2024}/\mathrm{MWh}$ | PPA cost-based pricing calculation based on following values. |
| | CAPEX | 3400 | $€_{2024}/\mathrm{kW}$ | [1] |
| | OPEX fix | 39 | $€_{2024}/(\mathrm{kW} \cdot \mathrm{a})$ | [1] |
| | OPEX var | 0.008 | $€_{2024}/\mathrm{kWh}$ | [1] |
| | Lifetime | 25 | $\mathrm{a}$ | [1] |
| | Annual production | 4406 | $\mathrm{MWh}/\mathrm{MW}$ | Based on used capacity factor time series, see Supplementary Note 2 |
| | Nominal weighted average cost of capital (WACC) | 7.9 | $\%$ | [1] |
| Wind onshore PPA (Germany / Spain) | Pay-as-produced price* | 70.4 / 73.8 | $€_{2024}/\mathrm{MWh}$ | PPA cost-based pricing calculation based on following values. |
| | CAPEX | 1900 / 1900 | $€_{2024}/\mathrm{kW}$ | [1] |
| | OPEX fix | 32 / 39 | $€_{2024}/(\mathrm{kW} \cdot \mathrm{a})$ | [1] |
| | OPEX var | 0.007 / 0.008 | $€_{2024}/\mathrm{kWh}$ | [1] |
| | Lifetime | 25 / 25 | $\mathrm{a}$ | [1] |
| | Annual production | 2802 / 3522 | $\mathrm{MWh}/\mathrm{MW}$ | Based on used capacity factor time series, see Supplementary Note 2 |
| | Nominal weighted average cost of capital (WACC) | 5.8 / 8.95 | $\%$ | [1] |
| Photovoltaic utility-scale PPA (Germany / Spain) | Pay-as-produced price* | 65.7 / 49.5 | $€_{2024}/\mathrm{MWh}$ | PPA cost-based pricing calculation based on following values. |
| | CAPEX | 900 / 900 | $€_{2024}/\mathrm{kW}$ | [1] |
| | OPEX fix | 13.3 / 13.3 | $€_{2024}/(\mathrm{kW} \cdot \mathrm{a})$ | [1] |
| | OPEX var | 0 / 0 | $€_{2024}/\mathrm{kWh}$ | [1] |
| | Lifetime | 30 / 30 | $\mathrm{a}$ | [1] |
| | Annual production | 1124 / 1777 | $\mathrm{MWh}/\mathrm{MW}$ | Based on used capacity factor time series, see Supplementary Note 2 |
| | Nominal weighted average cost of capital (WACC) | 5.4 / 7.3 | % | [1] |
| Grid (Germany / Spain) | Electricity price (reference assumption) | 93.5 / 81.3 | $€_{2024}/MWh$ | Energy and supply – Consumption 150 000 MWh and over – band IG – 2024 value [2] |
| | Average emission intensity of grid electricity | 298 / 129 | $kgCO_2e/MWh$ | 2024 value [3] |
| PEM electrolyser (@30bar output) | CAPEX | 1292.81 | $€_{2024}/kW$ | [4] |
| | OPEX fix | 20.12 | $€_{2024}/(kW \cdot a)$ | [4] |
| | Depreciation time | 15 | a | [4] |
| | Nominal weighted average cost of capital (WACC) | 8.9 | % | Calculated from real WACC value in [4], with expected inflation rate of 1.8 % |
| | Specific energy consumption at nominal load inclusive peripherals | 52.5 | $kWh/kgH_2$ | [4] |
| | Specific energy consumption peripherals | 3.33 | $kWh/kgH_2$ | [4] |
| | Decrease of specific energy consumption of stack at part load | 1 | % per 10% load reduction | [4] |
| | Specific water consumption | 14 | $kgH_2O/kgH_2$ | [4] |
| | Water cost | 3.725 | $€_{2024}/m^3H_2O$ | [4] |
| Compressor (@30bar input & 350bar/80bar**/80bar output - pressure tank/salt cavern (pipeline for transport)/pipeline for transport) | CAPEX | 4558.69 | $€_{2024}/kW$ | [4] |
| | OPEX fix | 4 | % of CAPEX/a | [4] |
| | Depreciation time | 15 | a | [4] |
| | Nominal weighted average cost of capital (WACC) | 9 | % | Calculated from real WACC value in [4], with expected inflation rate of 2 % |
| | $H_2$ losses | 0.5 | % | [4] |

| | Specific energy consumption for isentropic compression | 1.287/ 0.397**/ 0.397 | $kWh/kgH_2$ | Own calculation according to [5], [6] |
|---|---|---|---|---|
| | Isentropic efficiency | 80 | % | [7] |
| | Mechanical efficiency | 90 | % | [7] |
| Hydrogen storage – Pressure gas tank (@300 bar) | CAPEX | 730.57 | $€_{2024}/kg$ | [4] |
| | OPEX fix | 2 | % of CAPEX/a | [4] |
| | Depreciation time | 25 | a | [4] |
| | Nominal weighted average cost of capital (WACC) | 8.9 | % | Calculated from real WACC value in [4], with expected inflation rate of 1.8 % |
| Hydrogen storage – Salt cavern* | CAPEX / Capacity fee* | 12.75** | $\frac{€_{2024}}{kg \cdot a}$ | own calculations** |
| | OPEX var / Usage Fee** | 0.36** | $\frac{€_{2024}}{kg \cdot a}$ | own calculations** |
| | Annuity factor | 1*** | | |

* Details on PPA price calculation see below.
** The salt cavern storage is not assumed to be constructed solely for the electrolyser project analysed here but to be a rentable storage bundle. The annual rental fee is thereby composed of a capacity and a usage fee, covering, besides the actual storage of hydrogen, compression and all other above ground measures. Since salt cavern storage capacities are already marketed in this way for natural gas storage, fees from German salt cavern storage operators were respectively modified for hydrogen storage [8]. In this case, the compressor unit is required to increase the gas pressure to the pressure level of a pipeline for transport.
*** Assumption made to fit the annual capacity fee into the economical model presented in equation (4) in the Methods in the main document.

**Details on PPA price calculation**

To calculate the PPA pay-as-produced prices, the so called annuity method [1] is used and in the following applied to the Wind Offshore option as an example. First the annuity factor $A_{\mathrm{WindOff}}$ is calculated as follows, where $r_{\mathrm{in,WindOff}}$ is the interest rate and $t_{\mathrm{dep,WindOff}}$ the depreciation time. Since we assume long-term PPA contracts with fixed prices rather than indexed prices, the interest rate here displays the nominal weighted average cost of capital.

$$A_{\mathrm{WindOff}} = \frac{r_{\mathrm{in,WindOff}} \cdot \left(1 + r_{\mathrm{in,WindOff}}\right)^{t_{\mathrm{dep,WindOff}}}}{\left(1 + r_{\mathrm{in,WindOff}}\right)^{t_{\mathrm{dep,WindOff}}} - 1} \tag{1}$$

Subsequently the PPA price $p_{\mathrm{PPA,WindOff}}$ is calculated by adding up the annual capital expenditures $C_{\mathrm{TAC,CAPEX,WindOff}}$ and the variable and fixed annual operational expenditures $C_{\mathrm{TAC,OPEXfix,WindOff}}$ and $C_{\mathrm{TAC,OPEXvar,WindOff}}$ and dividing the sum by the annual production of the wind turbine $P_{\mathrm{WindOff,a}}$.

$$p_{\mathrm{PPA,WindOff}} = \frac{C_{\mathrm{TAC,CAPEX,WindOff}} + C_{\mathrm{TAC,OPEXfix,WindOff}} + C_{\mathrm{TAC,OPEXvar,WindOff}}}{P_{\mathrm{WindOff,a}}} \quad (2)$$

The annual capital expenditures are calculated by multiplying the capital expenditures $C_{\mathrm{CAPEX,WindOff}}$ by the annuity factor $A_{\mathrm{WindOff}}$.

$$C_{\mathrm{TAC,CAPEX,WindOff}} = C_{\mathrm{CAPEX,WindOff}} \cdot A_{\mathrm{WindOff}} \quad (3)$$

The variable annual operational expenditures are calculated by multiplying the variable operational expenditures $C_{\mathrm{OPEXvar,WindOff}}$ by the annual production of the wind turbine $P_{\mathrm{WindOff,a}}$.

$$C_{\mathrm{TAC,OPEXvar,WindOff}} = C_{\mathrm{OPEXvar,WindOff}} \cdot P_{\mathrm{WindOff,a}} \quad (4)$$

Finally, the fixed annual operational expenditures are the fixed operational expenditures $\mathrm{C}_{\mathrm{OPEXfix,WindOff}}$.

$$C_{\mathrm{TAC,OPEXfix,WindOff}} = C_{\mathrm{OPEXfix,WindOff}} \quad (5)$$

# Supplementary Note 2: Capacity factor time series for PPAs

The capacity factor time series, used in equations (20-22) in the Methods in the main document to map the feed-in and inter-year fluctuations of the PPAs, were obtained from [9]–[11] for the years 2000-2024. In order to get representative time series for realistic PPAs, the data was collected for three different locations in Germany (DE) and two different locations in Spain (ES). Once the location of an existing offshore wind park, once the location of an existing onshore wind park and once the location of an existing utility-scale photovoltaic park in Germany. Once the location of an existing onshore wind park and once the location of an existing utility-scale photovoltaic park in Spain. All PPA options and their locations are depicted in Supplementary Figure 1.

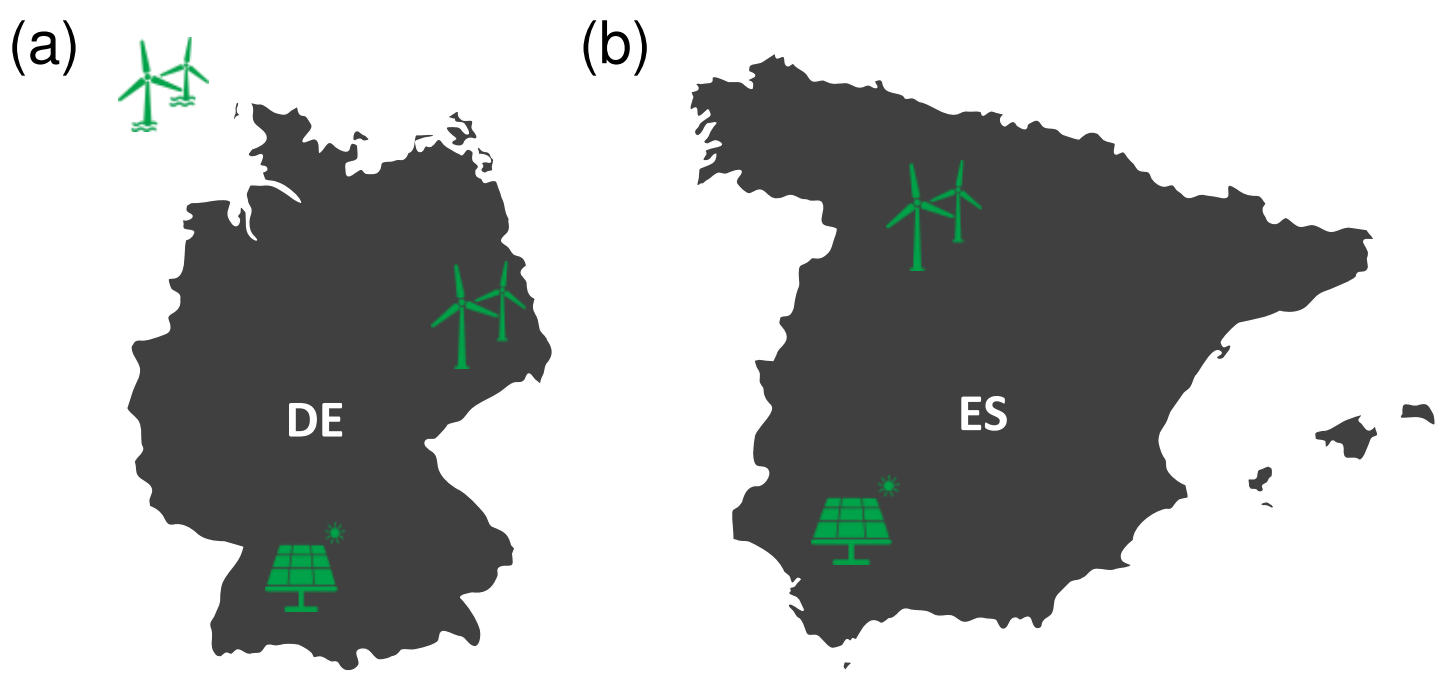

*Supplementary Figure 1: Analysed bidding zones and renewable PPA options and their geographical locations. (a), German bidding zone (DE) including wind offshore, wind onshore and utility-scale photovoltaic PPA options. (b), Spanish bidding zone (ES) including wind onshore and utility-scale photovoltaic PPA options.*

The chosen offshore wind park in Germany is *DanTysk* located in the North Sea. Since the wind park is mainly composed of Siemens SWT 120 3.6 wind turbines with a hub-height of 90 meters, the respective model was chosen to obtain the offshore wind capacity factor time series. The obtained capacity factor time series have an average capacity factor of 0.63. Lifetime capacity factors reported for this wind park are 0.5 [12], which is at the upper limit of expected capacity factors for new wind parks in the North Sea [1]. Such discrepancies are common in renewables feed-in data, which are based on reanalysis weather data [10]. Therefore, the capacity factor time series are corrected by multiplying them by the ratio between reported and their average capacity factor.

The chosen onshore wind park in Germany is *Bahren West Wind Park* located in the state of Brandenburg. Since the wind park is mainly composed of Vestas V150 5600 wind turbines with a hub-height of 166 meters, the respective model was chosen to obtain the onshore wind capacity factor time series. The obtained capacity factor time series have an average capacity factor of 0.32, which is within the expected range of newly built onshore wind farms in Germany [1], [13].

The chosen utility-scale photovoltaic park in Germany is *Solarpark Langenenslingen-Wilflingen* located in the state of Baden-Württemberg. Since the photovoltaic park is south oriented, a south orientation was chosen to obtain the photovoltaic capacity factor time series. The obtained capacity factor time series have an average capacity factor of 0.16. Expected capacity factors reported by the park operator are 0.128 [14], which is in the range of expected capacity factors in the region[1]. Therefore, the capacity factor time series are corrected as described above.

The chosen onshore wind park in Spain is *Andella Wind Farm* located in Valladolid Province. Since the wind park is mainly composed of Siemens Gamesa SG 5.0-145 wind turbines with a hub-height of 127.5 meters, the respective model was chosen to obtain the onshore wind capacity factor time series. The obtained capacity factor time series have an average capacity factor of 0.4, which is within the expected range of newly built onshore wind farms in Spain [1], [13], [15].

The chosen utility-scale photovoltaic park in Spain is *Guillena solar farm* located in Andalusia. Since the photovoltaic park is south oriented, a south orientation was chosen to obtain the photovoltaic capacity factor time series. The obtained capacity factor time series have an average capacity factor of 0.2, which is within the range of capacity factors expected in the region [1], [16].

**Choice of reference weather year for reference design optimisation**

The reference weather year and the respective capacity factor time series for the reference design optimisation described in the Methods and in Figure 2 (a) in the main document are chosen by finding the weather year with the lowest difference in average capacity factor compared to the average of all 25 analysed years. Therefore, the difference in capacity factor between the average value of each year and the average value of all 25 years is calculated for each PPA option. Subsequently, for each PPA option the weather years are rated with values 1 to 25 according to their absolute difference to the average capacity factor of all years. Thereby, the years with the lowest difference get the lowest rating. Finally, for each year, the ratings of each PPA option are summed up and the reference weather year is the year with the lowest rating sum. Different approaches could also be suitable to find a reference weather year. Supplementary Figure 2 shows the average capacity factors of all capacity factor time series for all PPA options and both bidding zones and the chosen reference year.

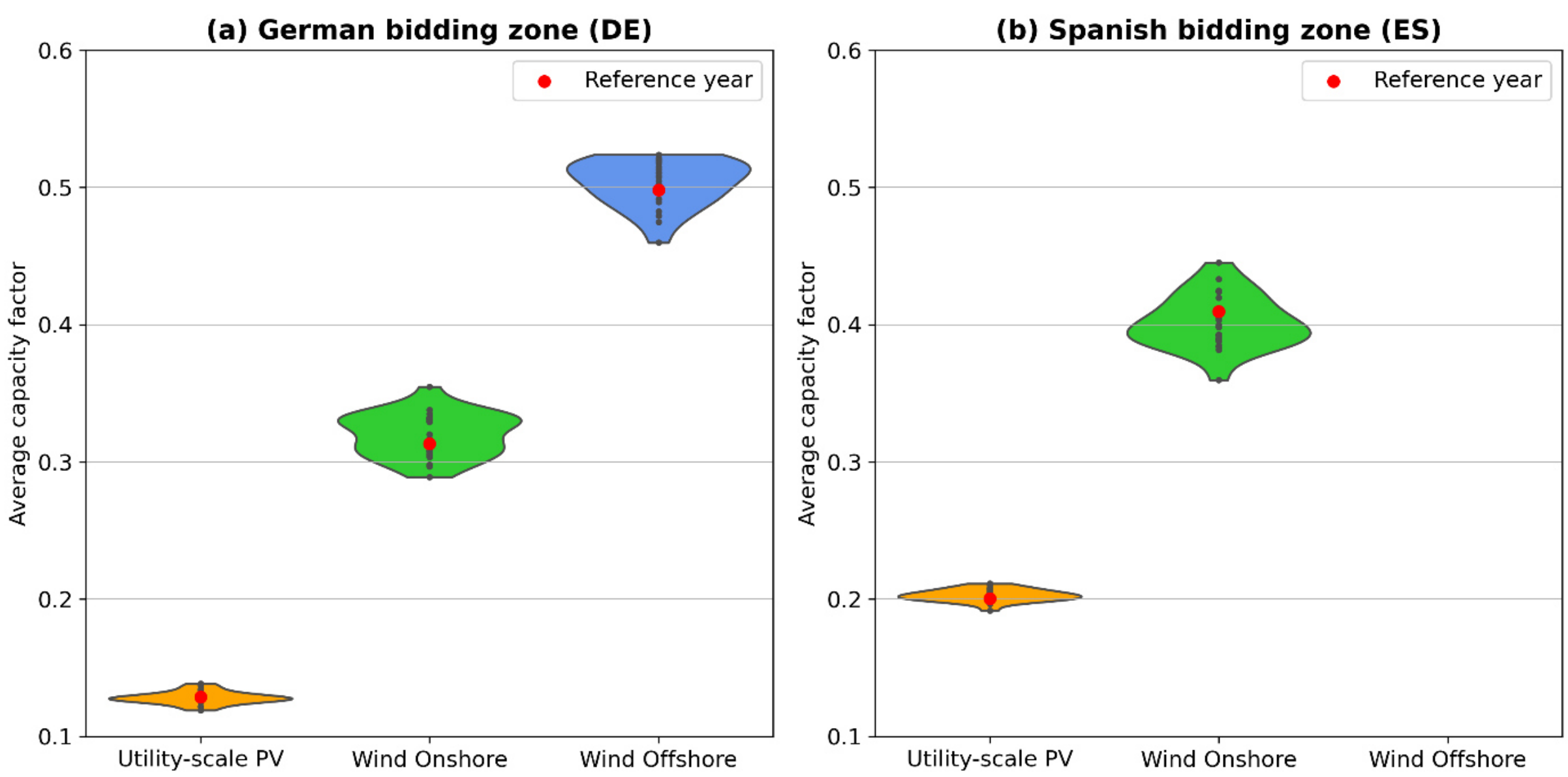

*Supplementary Figure 2: Average capacity factors of capacity factor time series for all PPA options and bidding zones. (a), German bidding zone (DE). (b), Spanish bidding zone (ES). Red dot shows reference year chosen for reference design optimisation.*

# Supplementary Note 3: Additional results *Relaxed rules for temporal correlation*

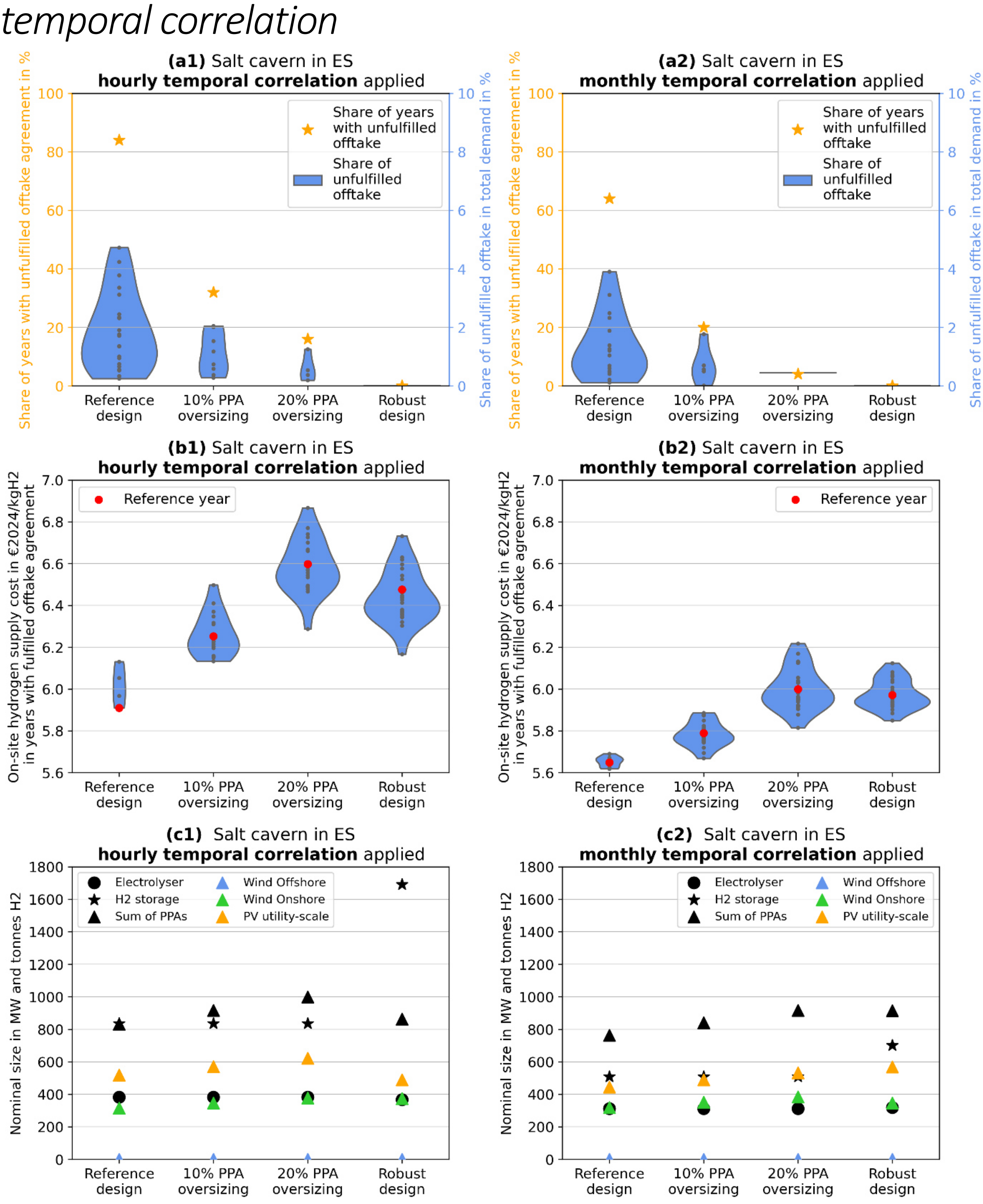

*Supplementary Figure 3: Comparison of optimisation results between the standard configuration of the considered green hydrogen production system with hourly (a1-c1) and monthly temporal correlation (a2-c2) for all analysed design paradigms. (a1) and (a2), yellow y-axis: Share of years with unfulfilled offtake agreement out of all analysed years; blue y-axis: Share of total demand needed to meet unfulfilled offtake agreement in years with unfulfilled offtake agreement. (b1) and (b2), On-site hydrogen supply cost in years with fulfilled offtake agreement. (c1) and (c2), Nominal sizes of electrolyser, storage and PPAs. The storage option is a salt cavern storage bundle. The bidding zone is the Spanish (ES) bidding zone. The hydrogen offtake characteristic is flat. The average electricity price for non-household consumers in 2024 in Spain with a consumption of 150 GWh or above is used as average grid electricity price for additional grid electricity purchase. [27]*

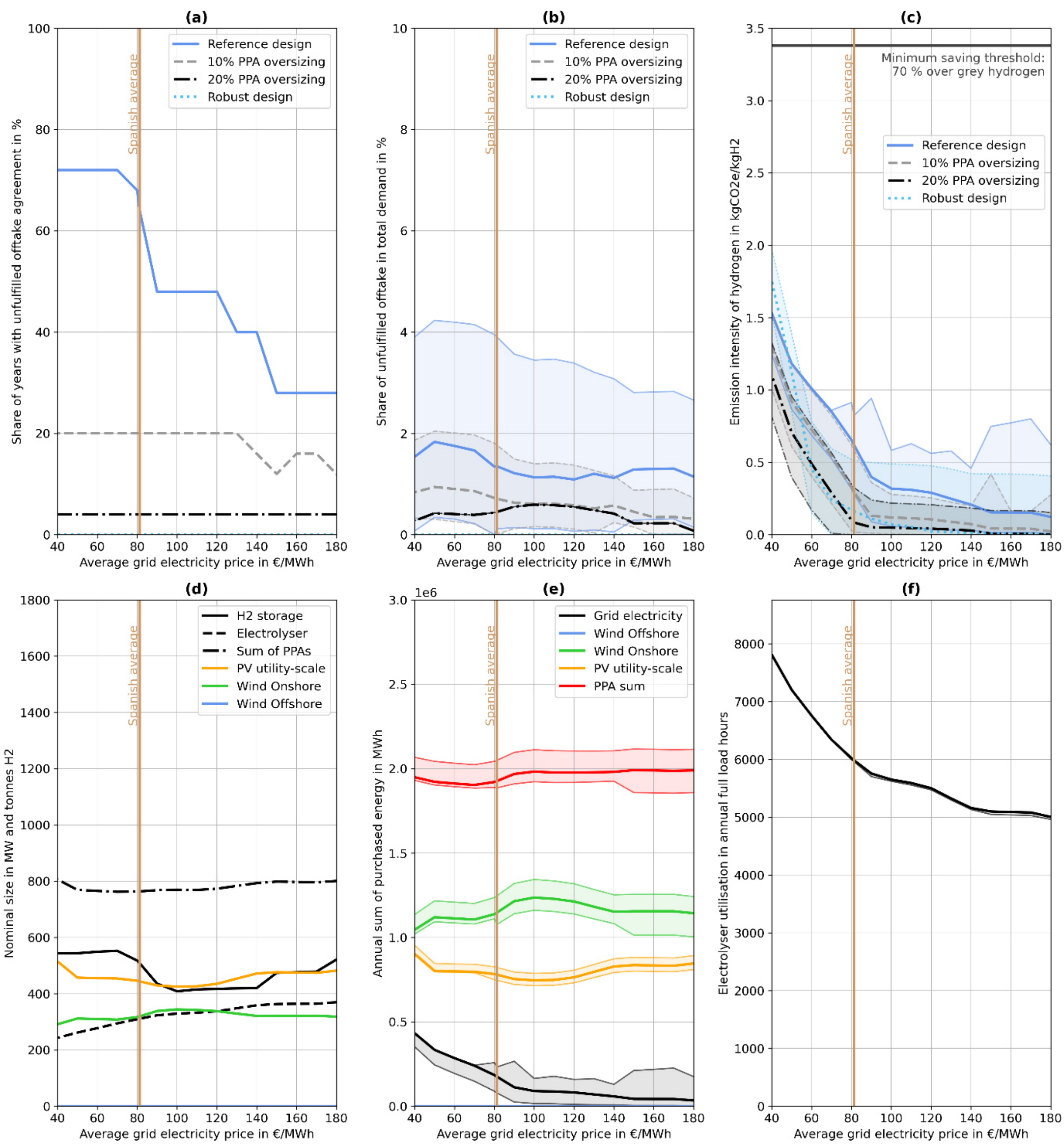

*Supplementary Figure 4: Optimisation results for the modified green hydrogen production system with monthly instead of hourly temporal correlation. (a), Share of years with unfulfilled offtake agreement out of all analysed years as a function of the average grid electricity price. (b), Share of total demand needed to meet unfulfilled offtake agreement in years with unfulfilled offtake agreement as a function of the average grid electricity price. (c), Emission intensity of the produced hydrogen as a function of the average grid electricity price in years with a fulfilled offtake agreement. (d), Nominal sizes of electrolyser, storage and PPAs under the reference design paradigm as a function of the average grid electricity price. (d), Annual sum of purchased power from the different power purchase options in the reference design paradigm as a function of the average grid electricity price. (e), Utilisation of the electrolyser in full load hours under the reference design paradigm as a function of the average grid electricity price. Thick lines within coloured areas in (b) show mean, thin lines on area boundaries show maximum and minimum values. Thick lines within coloured areas in (c), (e) and (f) show reference-year values, thin lines on area boundaries show minimum and maximum values. The storage option is a salt cavern storage bundle. The bidding zone is Spain (ES ). The hydrogen offtake characteristic is flat. (a)-(f) additionally show the average electricity price in Spain for non-household consumers in 2024 with a consumption of 150 GWh or above as a refer-ence [21]. In addition, (c) displays the minimum emission intensity saving threshold set for green hydrogen, which is 70 % over its fossil comparator grey hydrogen [33].*

# Supplementary Note 4: Additional results *Increased offtake flexibility*

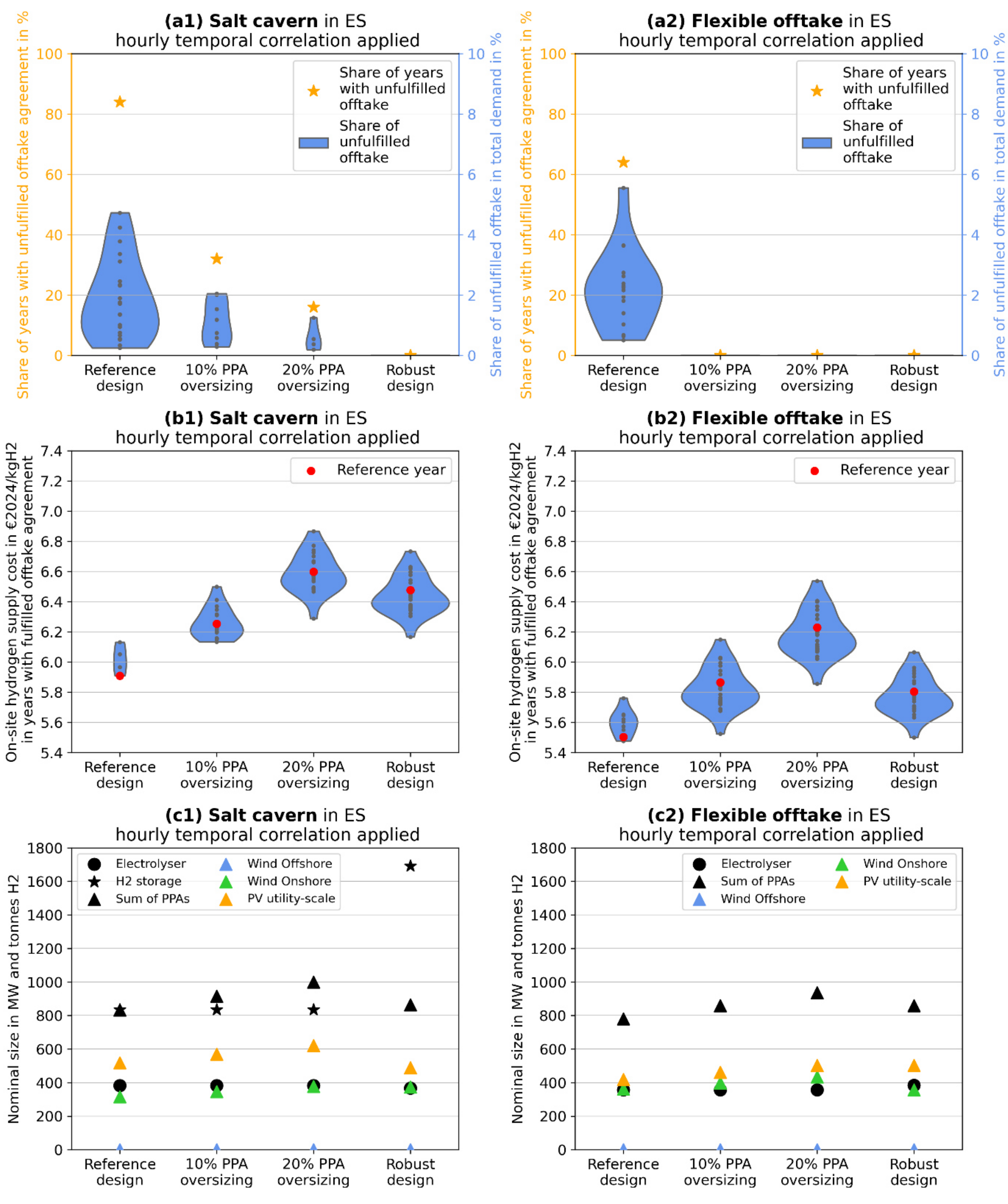

*Supplementary Figure 5: Comparison of optimisation results between the standard configuration of the considered green hydrogen production system with a salt cavern storage option and a flat offtake characteristic (a1-c1) and a time flexible offtake where no hydrogen storage is needed (a2-c2) for all analysed design paradigms. (a1) and (a2), yellow y-axis: Share of years with unfulfilled offtake agreement out of all analysed years; blue y-axis: Share of total demand needed to meet unfulfilled offtake agreement in years with unfulfilled offtake agreement. (b1) and (b2), On-site hydrogen supply cost in years with fulfilled offtake agreement. (c1) and (c2), Nominal sizes of electrolyser, storage and PPAs. The bidding zone is the Spanish (ES) bidding zone.*

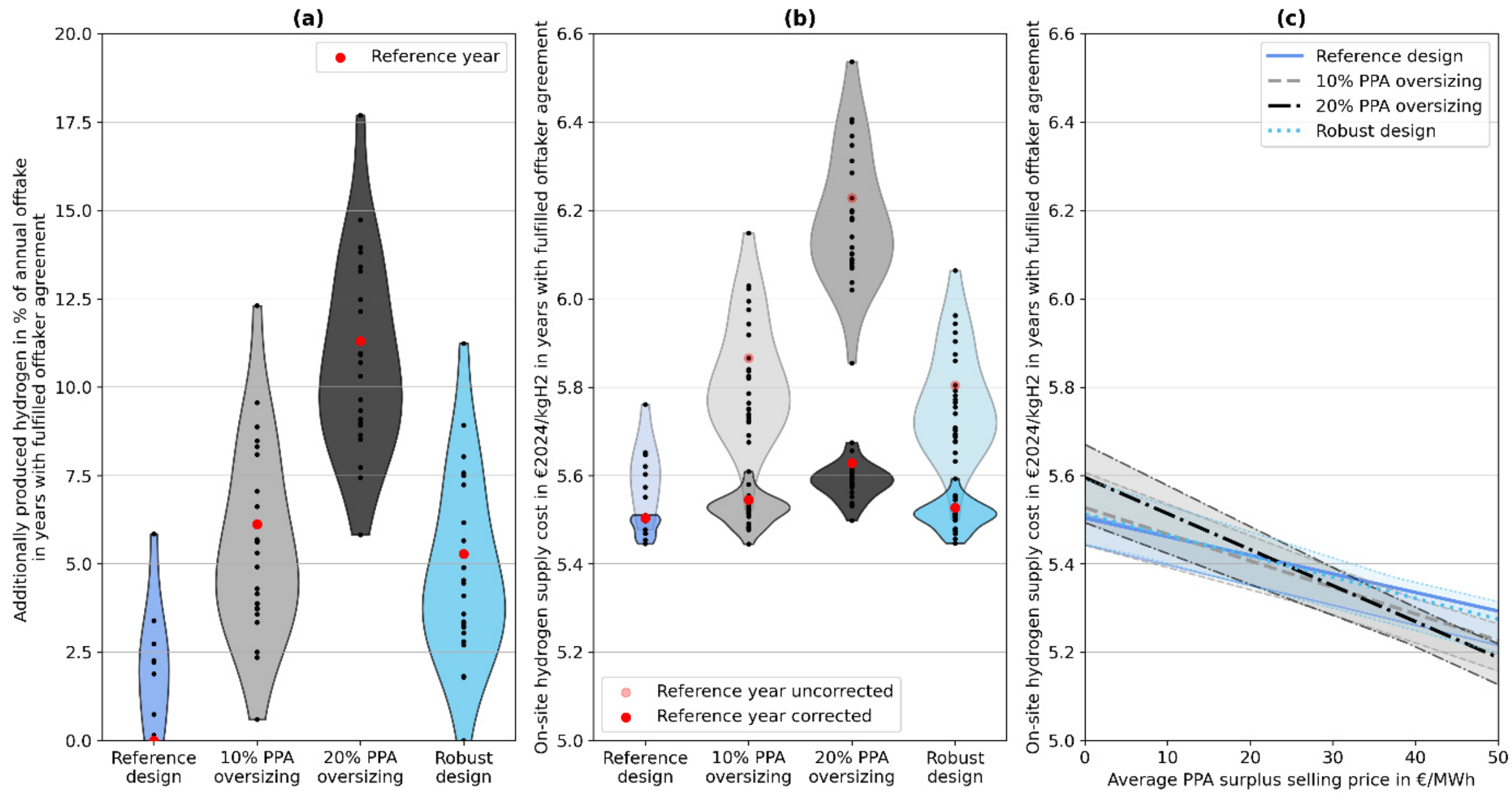

*Supplementary Figure 6: Optimisation results for the modified green hydrogen production system with a time flexible offtake where no hydrogen storage is needed under all analysed design paradigms with the additional use of surplus PPA power for combined green hydrogen production and PPA surplus selling. (a), Additionally produced hydrogen from PPA surplus use in % of annual offtake in years with fulfilled offtake agreement. (b), Dark coloured violin plots: On-site hydrogen supply cost in years with fulfilled offtake agreement corrected by additionally supplied hydrogen from surplus use. Pale coloured violin plots: On-site hydrogen supply cost in years with fulfilled offtake agreement uncorrected. (c), On-site hydrogen supply cost in years with fulfilled offtake agreement corrected by additional hydrogen production and PPA surplus selling revenues as a function of the PPA surplus selling price. Thick lines within coloured areas in (c) show reference-year values, thin lines on area boundaries show maximum and minimum values. The storage option is a salt cavern storage bundle. The bidding zone is Spain (ES). The temporal correlation condition is hourly. The hydrogen offtake characteristic is flat plus additional hydrogen produced from PPA surplus use on top.*